\documentclass[aps, prapplied, reprint, superscriptaddress]{revtex4-2}

\usepackage{graphicx}
\usepackage{amsmath}
\usepackage{amssymb}
\usepackage{siunitx}
\usepackage{booktabs}
\usepackage{hyperref}
\usepackage{xcolor}

\hypersetup{
    colorlinks=true,
    linkcolor=blue,
    citecolor=blue,
    urlcolor=blue
}

\begin{document}

\title{\textbf{Beyond binary at submicron dimensions: crossed-ellipse MTJ free layers as multi-state cells for spintronic crossbars}}

\author{Arup Ghosh}
\email{arupghoshphy@gmail.com}
\affiliation{Department of Physics, Institute of Nanotechnology and Advanced
Materials, Bar-Ilan University, Ramat Gan 5290002, Israel}

\author{Ariel Zaig}
\affiliation{Department of Physics, Institute of Nanotechnology and Advanced
Materials, Bar-Ilan University, Ramat Gan 5290002, Israel}

\author{Colin Ducarme}
\affiliation{Universit\'e catholique de Louvain, Institute of Condensed Matter
and Nanosciences, Louvain-la-Neuve 1348, Belgium}

\author{Thomas Copp\'ee}
\affiliation{Universit\'e catholique de Louvain, Institute of Condensed Matter
and Nanosciences, Louvain-la-Neuve 1348, Belgium}

\author{Flavio Abreu \surname{Araujo}}
\affiliation{Universit\'e catholique de Louvain, Institute of Condensed Matter
and Nanosciences, Louvain-la-Neuve 1348, Belgium}

\author{Lior Klein}
\email{lior.klein@biu.ac.il}
\affiliation{Department of Physics, Institute of Nanotechnology and Advanced
Materials, Bar-Ilan University, Ramat Gan 5290002, Israel}

\begin{abstract}
Spintronic crossbars are among the most promising hardware substrates for the energy-efficient neuromorphic computation that modern AI workloads increasingly demand, yet each magnetic tunnel junction (MTJ) in such an array is currently binary, which places a hard ceiling on the synaptic precision and information density that a single crosspoint can carry. Lifting this ceiling requires MTJs that host multiple readable resistance states while remaining small enough for dense integration. MTJs in which the soft magnetic layer is formed from two crossed ellipses have been shown to support four field-free, spin--orbit-torque (SOT)-switched remanent states, so far only at micrometer dimensions. Whether more states can be stabilized, and whether the structure can be shrunk to submicron dimensions, have not been systematically studied. Here, we address both questions using MuMax3 simulations of permalloy (Py) crossed-ellipse free layers, treated as a well-characterized micromagnetic model system; transferability of the geometry-driven trends to CoFeB free layers used in practical MTJ stacks is also tested. At fixed \(8:1\) aspect ratio, shrinking the device from \(16~\si{\micro m}\times2~\si{\micro m}\) to \(80~\si{nm}\times10~\si{nm}\) lowers the absolute switching current, which is favorable for array drivers, but raises the switching field from \(\sim\SI{11}{Oe}\) to \(\sim\SI{301}{Oe}\) and increases the SOT current density, defining a practical lower-size bound. At fixed major axis \(a=\SI{1.6}{\micro m}\), aspect-ratio tuning gives a low-field window for the ordinary four-state branch, while lower-aspect-ratio geometries stabilize additional remanent states with lower switching fields and current densities. The \(1.6~\si{\micro m}\times0.8~\si{\micro m}\) geometry hosts twelve accessible remanent plateaus in its angle-resolved planar Hall response, and state-count mapping shows that the number of resolved plateaus depends on both aspect ratio and absolute lateral size. Projection-based MTJ estimates show that these plateaus can be mapped into multiple electrical levels, while minimum-energy-path calculations show that the twelve accessible configurations are not all thermally independent. In particular, the closest canted states are connected by very small barriers, and independent MuMax+ calculations indicate that lower-barrier multistep escape pathways can also connect nominally more distant states. The present geometry should therefore be interpreted as a multistate free-layer landscape whose nonvolatile state count requires further optimization of geometry and material parameters.
\end{abstract}
\maketitle
\vspace{0.5cm}


\section{Introduction}
\vspace{-0.25cm}
The increasing computational load associated with artificial intelligence has renewed interest in hardware architectures that can perform memory and computation in the same physical array. Crossbar networks are attractive for this purpose because conductance values at the crosspoints can directly encode synaptic weights and support highly parallel multiply--accumulate operations \cite{jungMagnetoresistiveCrossbar}. Spintronic crossbars are a promising member of this broader device family because magnetic tunnel junctions (MTJs) provide nonvolatile, reproducible, and electrically readable resistance states \cite{apalkovReview,kentWorledge,marrowsSpintronics2024}. However, a conventional MTJ is intrinsically binary, so a single crosspoint normally stores only two conductance levels. Increasing the number of stable resistance states per cell is therefore a direct route toward improving information density and synaptic precision in spintronic crossbars \cite{multistateAEM,dwMTJSynapseNatComm2024,loneNanoscale2024,kumarNanoscaleHorizons2024}.

One way to go beyond binary operation is to replace a purely perpendicular-anisotropy free layer with a geometry-engineered in-plane magnetic element. In-plane magnetic anisotropy can use shape anisotropy, demagnetizing fields, and magnetostatic coupling to define several stable magnetization directions \cite{osborn,aharoni,stonerwohlfarth,brown,cowburn}. Crossed-ellipse magnetic structures are especially attractive because the two arms and their central overlap region can create a multi-minimum energy landscape. A previous MTJ consisting of a single elliptical reference layer and a two-crossed-ellipse soft layer demonstrated four distinct resistance states and field-free SOT switching through an underlying Ta layer \cite{dasAPL}. That experiment established the crossed-ellipse geometry as a viable multistate MTJ concept, but the demonstrated device remained at micrometer dimensions and supported only four states.

Other shape-engineered magnetic elements, including multi-arm artificial-spin-ice vertices, Y-shaped domain-wall devices, and other non-elliptical nanomagnets, have also been explored as routes to geometry-controlled magnetic states and reconfigurable magnetoresistive or logic response \cite{skjaervoASIReview,hrabecYDevice,shapeASI2025}. Compared with these alternatives, the crossed-ellipse geometry is attractive because it preserves a simple two-arm layout, has already been experimentally integrated into a four-state MTJ, and allows the state count to be tuned by aspect ratio and absolute size \cite{dasAPL}.

For crossbar integration, two questions then become central. First, how small can the crossed-ellipse cell be made before its remanent states or switching process become unreliable? Second, can geometry be used not only to preserve the four-state operation but also to stabilize additional states? Miniaturization improves areal density and can reduce the absolute write current, but patterned nanomagnets do not scale trivially: exchange, demagnetizing fields, edge roughness, and effective magnetic thickness can all modify the reversal pathway and the switching field \((H_s)\) \cite{ross,permalloyNanodots,cylindricalMTJ2025}. Current-driven switching adds another constraint because the relevant device stress is often set by the switching current density \((J_s)\), not only by the absolute switching current \((I_s)\) \cite{liuTa,garello,krizakovaReview}.

SOT switching in heavy-metal/ferromagnet bilayers provides an efficient write mechanism for such multistate devices. A charge current in the heavy-metal layer generates a transverse spin current through the spin Hall effect, exerting torque on the adjacent ferromagnet \cite{liuPRL,liuTa,mironNature,manchonReview}. This three-terminal geometry is attractive for MTJ arrays because read and write paths can be separated, improving endurance and write selectivity \cite{garello,wangNatElectron,krizakovaReview}. The sign and magnitude of the spin Hall angle depend on the heavy metal, so materials such as Ta and Pt can modify the switching efficiency and the relative role of SOT and Oersted-field contributions \cite{liuTa,paiAPL,manchonReview}.

In this work, micromagnetic simulations are used to define the geometry-dependent design window for crossed-ellipse MTJ free layers. Permalloy (Py) is used as a well-characterized model material, while supplementary CoFeB calculations test whether the main trends persist for an MTJ-relevant ferromagnet. The study follows two complementary design axes relevant to multistate spintronic crossbar cells. First, the device size is reduced at fixed aspect ratio \(a:b=8:1\), from \(16~\si{\micro m}\times2~\si{\micro m}\) to \(80~\si{nm}\times10~\si{nm}\), to determine the scaling of the switching field \(H_s\), absolute switching current \(I_s\), and SOT switching current density \(J_s\). Second, the aspect ratio is varied at fixed major axis \(a=\SI{1.6}{\micro m}\) to identify the low-field four-state operating branch, relate the high-aspect-ratio switching trend to an effective crossed-ellipse anisotropy field \(H_k\), and determine when additional remanent states appear. Supplementary fixed-major-axis calculations, state-count mapping, projected MTJ resistance estimates, and representative string-method barrier calculations are then used to separate aspect-ratio effects from absolute-size effects and to distinguish accessible field-off plateaus from thermally robust memory basins.

\vspace{-0.25cm}
\section{Simulation Method and Device Geometry}
\vspace{-0.25cm}
Micromagnetic simulations were performed using MuMax3 \cite{mumax3}. The MuMax input files were generated using a Python script, which defined the simulation geometry, material parameters, measurement angles, magnetic field sequence, current-pulse sequence, and output tables. The structure consisted of two crossed Py ellipses, where the second ellipse was obtained by rotating the base ellipse by \(90^\circ\). The total simulated geometry was defined as the union of the two ellipses.

For the representative \(8:1\) geometry, the major and minor axes were \(a=\SI{1.6}{\micro m}\) and \(b=\SI{0.2}{\micro m}\). The Py thickness was fixed at \(t_{\mathrm{Py}}=\SI{2}{nm}\). A dead-layer thickness of \(t_{\mathrm{dead}}=\SI{0.3}{nm}\) was included as a phenomenological correction to the active Py thickness. This value is used here as a conservative interface correction, motivated by reports that Ta seed and cap layers can reduce the effective magnetic thickness of ultrathin NiFe films \cite{kowalewskiTaPyDeadLayer}. A no-dead-layer comparison is provided in the Supplementary Information to show how this assumption affects the switching-field scale.

The saturation magnetization was scaled according to the effective magnetic thickness using \(M_s=800\times10^3(t_{\mathrm{eff}}/t_{\mathrm{Py}})~\mathrm{A\,m^{-1}}\), giving \(M_s=6.8\times10^5~\mathrm{A\,m^{-1}}\). The intrinsic Py exchange length was fixed at \(\ell_{\mathrm{ex}}=\SI{5.7}{nm}\). After applying the dead-layer correction to \(M_s\), the exchange stiffness was recalculated using \(A_{\mathrm{ex}}=\mu_0 M_s^2\ell_{\mathrm{ex}}^2/2\). For \(M_s=6.8\times10^5~\mathrm{A\,m^{-1}}\), this gives \(A_{\mathrm{ex}}\approx9.45\times10^{-12}~\mathrm{J\,m^{-1}}\). This procedure treats the dead-layer correction as an effective-volume correction while preserving the intrinsic Py exchange length. The Gilbert damping constant was set to \(\alpha=0.01\), a commonly used value for Py micromagnetic simulations and thin-film magnetization dynamics; this parameter mainly affects the relaxation and dynamic switching timescale, while the quasi-static field-driven trends are governed primarily by the magnetic energy landscape \cite{gilbertPyDynamics,pyDampingTemperature}.
The computational cell size was chosen according to the lateral device size.
For the initial micrometer-scale and intermediate-size geometry scans, the
in-plane cell size was \(\Delta x=\Delta y=\SI{5}{nm}\). For devices with
lateral dimensions below \(\SI{480}{nm}\), a finer \(\SI{3}{nm}\) in-plane
cell size was used, and for devices with lateral dimensions of
\(\SI{160}{nm}\) and below, the in-plane cell size was further reduced to
\(\SI{2}{nm}\). The out-of-plane cell size followed the active magnetic
thickness used in the corresponding simulation. Edge smoothing was enabled
with \(\mathrm{EdgeSmooth}=8\).

Because the low-aspect-ratio multistate response is one of the central results
of this work, the \(a=\SI{1.6}{\micro m}\),
\(b=\SI{0.8}{\micro m}\) geometry was additionally recalculated using a refined
\(\Delta x=\Delta y=\SI{2}{nm}\) in-plane discretization. The twelve-plateau
field-off response was reproduced with the finer mesh, confirming that the
multistate landscape is not a consequence of the
\(\SI{5}{nm}\) discretization used in the initial geometry scan. Independent
MuMax+ convergence calculations similarly showed a clear convergence trend
with decreasing in-plane cell size and identified an in-plane cell size of
approximately \(\SI{2}{nm}\) as a suitable accuracy--cost compromise for the
tested geometries (Fig.~S19(a-d)).

The simulation geometry was divided into regions corresponding to the two ellipses and the central overlap region. This allowed the initial magnetization in each part of the device to be assigned independently. For the crossed-ellipse remanent-state simulations, the ellipses were initialized along their respective easy axes (\(0^\circ\) and \(90^\circ\)) and the overlap region was initialized along \(45^\circ\). The initialized magnetic structure was obtained by relaxing the geometry under zero-field conditions. The representative four primary remanent states of such a device with aspect ratio \(a:b=8:1\) are shown in Fig.~\ref{fig:remnant_states_scaling}(a).

The applied magnetic field \((H)\) in Oe was swept along selected in-plane directions. The corresponding MuMax field components were defined as \(B_x=H\times10^{-4}\cos\theta\) and \(B_y=H\times10^{-4}\sin\theta\), where \(\theta\) is the in-plane field angle. The simulations considered field directions such as \(45^\circ\), \(135^\circ\), \(225^\circ\), and \(315^\circ\), consistent with the crossed-ellipse remanent-state configurations. For each field condition, the magnetization was relaxed, and then the field was removed to obtain the remanent state.

For SOT switching simulations, a Ta/Py bilayer geometry was considered, with a \(\SI{5}{nm}\) Ta heavy-metal layer placed underneath the Py free layer. The current was applied along the horizontal ellipse through the Ta layer, and the SOT generated by the spin Hall effect was applied to the Py magnetization. The SOT simulations were performed at \(H=0\). For current partitioning, representative thin-film resistivities were used: \(\rho_{\mathrm{FM}}=1.25\times10^{-6}~\Omega\mathrm{m}\) for Py and \(\rho_{\mathrm{HM}}=1.8\times10^{-6}~\Omega\mathrm{m}\) for Ta. The Py value was treated as an effective ultrathin-film resistivity, since NiFe resistivity is strongly affected by thickness and grain-boundary scattering \cite{jaoulPyResistivity,boonaPyThermalElectrical}. The Ta value is consistent with the high resistivity commonly reported for \(\beta\)-Ta thin films \cite{cataniaBetaTa,readAltmanBetaTa}. The corresponding heavy-metal current density was then calculated from the Ta/Py parallel conduction path for each applied current pulse. The spin Hall angle of Ta was taken as \(\theta_{\mathrm{SH}}=-0.15\). In MuMax3, the SOT was implemented using the Slonczewski torque term as an effective SOT-like torque, while Zhang--Li torque was disabled.

The current-driven switching was simulated using rectangular current pulses with abrupt rise and fall times. For each trial current amplitude, the current was applied for \(\SI{0.9}{ns}\), followed by a \(\SI{5}{ns}\) current-off relaxation step to determine the final remanent state. Switching was identified when the post-pulse relaxed state corresponded to the target remanent configuration. Because the threshold current can depend on pulse duration and waveform, the reported \(I_s\) and \(J_s\) values should be interpreted for this pulse protocol.

The Oersted field associated with the heavy-metal current was estimated as \(B_{\mathrm{Oe}}=\mu_0 J_{\mathrm{HM}}t_{\mathrm{HM}}/2\) and included in the driven region during the current pulse, where \(J_{\mathrm{HM}}\) and \(t_{\mathrm{HM}}\) are respectively the current density and thickness of the underlying Ta layer. This expression treats the heavy-metal layer as a locally uniform current sheet and therefore neglects lateral nonuniformity of the current distribution and edge fields. The approximation captures the leading Oersted-field scale for comparing geometries, but it may become less accurate for the smallest devices. Therefore, the calculated SOT thresholds should be interpreted as comparative geometry-dependent trends rather than exact device-level switching currents.
\vspace{-0.25cm}
\section{Simulation Results}
\vspace{-0.25cm}
\subsection{Device-size scaling at fixed aspect ratio}

\begin{figure}[t!]
    \centering
    \includegraphics[width=\columnwidth, trim=2cm 0.5cm 2cm 0cm, clip]{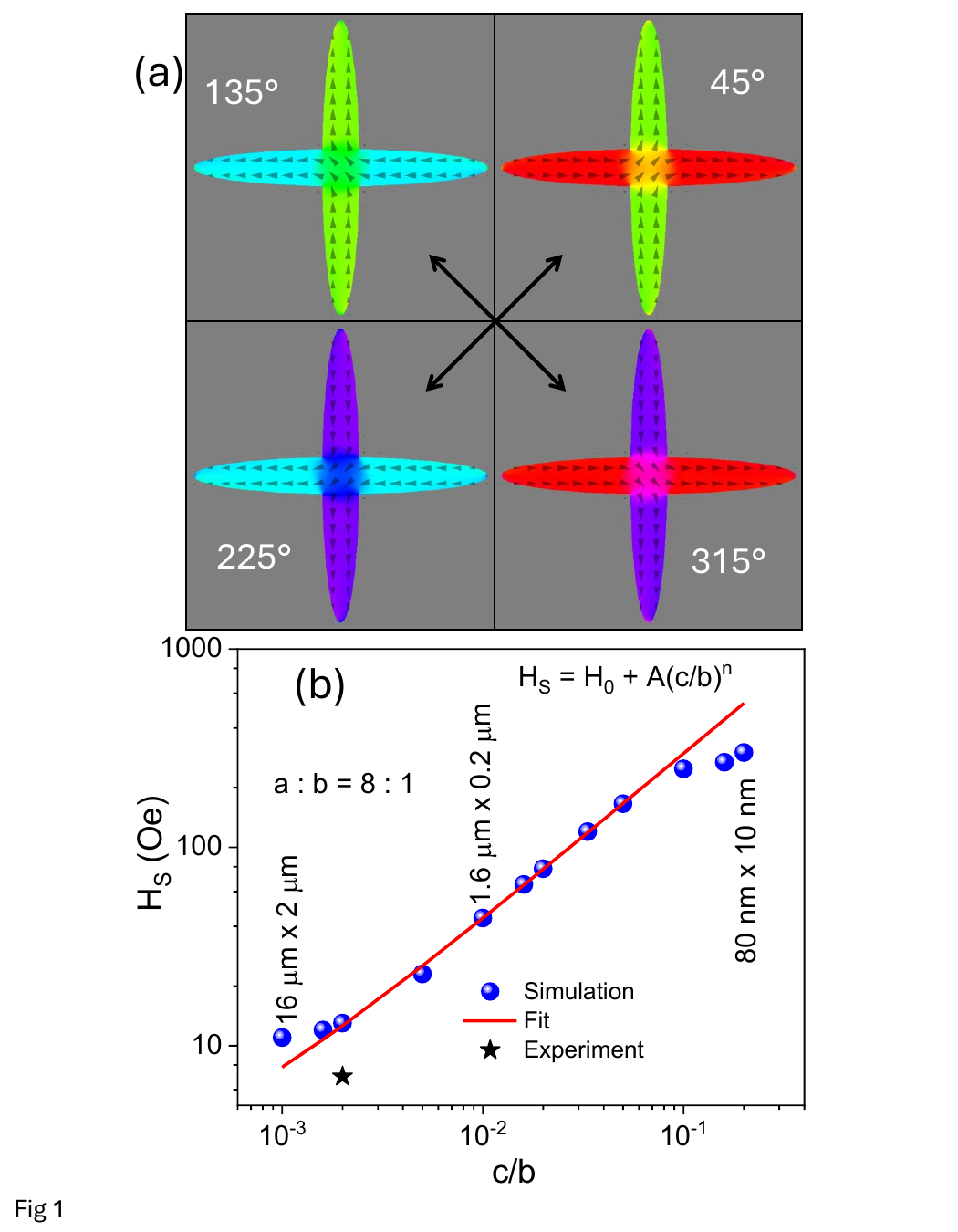}
    \caption{
    Remanent-state configurations and size-dependent switching behavior of the crossed-ellipse device. Panel (a) shows the representative four remanent states for the \(8:1\) geometry, corresponding to the diagonal directions \(45^\circ\), \(135^\circ\), \(225^\circ\), and \(315^\circ\). Panel (b) shows the switching-field scaling with normalized thickness-to-width ratio \(c/b\) at fixed aspect ratio \(a:b=8:1\), together with a phenomenological fit. The black star symbol indicates the experimental switching field reported for the \(8~\si{\micro m}\times1~\si{\micro m}\) four-state MTJ in Ref.~\cite{dasAPL}; the simulation overestimates this value by about a factor of two, consistent with the simplified isolated-free-layer model used here.
    }
    \label{fig:remnant_states_scaling}
    \vspace{-0.5cm}
\end{figure}

The first set of simulations considered size reduction at fixed aspect ratio \(a:b=8:1\). In this series, both \(a\) and \(b\) were reduced while maintaining the same in-plane shape. The thickness remained fixed at \(c=\SI{2}{nm}\). Therefore, reducing the lateral size increased the normalized thickness-to-width ratio \(c/b\). Selected device dimensions and the corresponding \(H_s\), \(I_s\), and \(J_s\) values are summarized in Table~\ref{tab:size_scaling}. The switching-field scaling is shown in Fig.~\ref{fig:remnant_states_scaling}(b), while the corresponding \(I_s\) and current-density trends are shown in Fig.~\ref{fig:current_scaling}.

\begin{table}[t!]
\centering
\renewcommand{\arraystretch}{1.2}
\setlength{\tabcolsep}{6pt}
\small
\caption{
Switching field \((H_s)\), absolute switching current \((I_s)\), and SOT switching current density \((J_s)\) as a function of device size and normalized thickness-to-width ratio \(c/b\) for crossed Py ellipses with fixed aspect ratio \(a:b=8:1\) and fixed Py thickness \(c=\SI{2}{nm}\). The current-driven values correspond to the Ta/Py bilayer switching protocol, using a \(\SI{0.9}{ns}\) current pulse followed by \(\SI{5}{ns}\) relaxation. The reported \(I_s\) and \(J_s\) values are therefore protocol-specific switching thresholds rather than universal material constants.
}
\label{tab:size_scaling}
\begin{tabular}{ccccc}
\toprule
Device size & \(c/b\) & \(H_s\) & \(I_s\) & \(J_s\) \\
 &  & (Oe) & (mA) & \((10^{11}~\mathrm{A\,m^{-2}})\) \\
\midrule
\(16~\si{\micro m}\times2~\si{\micro m}\) & 0.001 & 10.8 & 3.50 & 2.50 \\
\(8~\si{\micro m}\times1~\si{\micro m}\) & 0.002 & 13.1 & 2.10 & 3.00 \\
\(1.6~\si{\micro m}\times0.2~\si{\micro m}\) & 0.010 & 44.2 & 0.88 & 6.29 \\
\(0.8~\si{\micro m}\times0.1~\si{\micro m}\) & 0.020 & 78.1 & 0.72 & 10.29 \\
\(160~\si{nm}\times20~\si{nm}\) & 0.100 & 248.7 & 0.42 & 30.00 \\
\(80~\si{nm}\times10~\si{nm}\) & 0.200 & 301.1 & 0.24 & 34.29 \\
\bottomrule
\end{tabular}
\end{table}

The switching field increases strongly as the device is reduced in size. The value of \(H_s\) is \(\sim\SI{11}{Oe}\) for the largest \(16~\si{\micro m}\times2~\si{\micro m}\) device, but increases to \(\sim\SI{301}{Oe}\) for the \(80~\si{nm}\times10~\si{nm}\) device. This shows that maintaining the same aspect ratio while reducing the absolute device size increases the magnetic switching requirement.

This increase can be understood in terms of stronger confinement and enhanced effective shape anisotropy as the width decreases. Results for devices smaller than \(80~\si{nm}\times10~\si{nm}\) are omitted because the magnetic states became unstable in this regime. This instability is likely related to the minor axis approaching only a few exchange lengths, where exchange-dominated finite-size effects become significant \cite{brown,ross,cowburn}. The smallest-device simulations used a refined in-plane cell size of \(\SI{2}{nm}\), so the \(80~\si{nm}\times10~\si{nm}\) geometry was resolved by approximately five cells across its minor axis. In the simulations, the Py exchange length was fixed at \(\ell_{\mathrm{ex}}=\SI{5.7}{nm}\) by scaling \(A_{\mathrm{ex}}\) consistently with the dead-layer-corrected \(M_s\). Thus, the \(80~\si{nm}\times10~\si{nm}\) device should be viewed as an indicative lower-size boundary for the present geometry and material parameters, rather than as a guaranteed practical crossbar-cell size.

The switching-field dependence on \(c/b\) was described using the phenomenological relation \(H_s=H_0+A(c/b)^n\), as shown in Fig.~\ref{fig:remnant_states_scaling}(b). Here, \(H_0\) is an effective offset field that represents the extrapolated switching-field baseline in the low-\(c/b\) limit, rather than a directly measured physical constant. The coefficient \(A\) describes the strength of the geometry-dependent contribution, while \(n\) describes the sensitivity of \(H_s\) to changes in \(c/b\). For the low-to-intermediate \(c/b\) range, \(0.001\leq c/b\leq0.020\), the fitted parameters were approximately \(H_0=\SI{1.8}{Oe}\), \(A=\SI{2.1e3}{Oe}\), and \(n=0.85\), with \(R^2=0.996\). The fit captures the main low-to-intermediate scaling trend, but deviations outside this range, especially for the smallest devices, indicate that it should be interpreted as a phenomenological scaling description rather than a universal reversal law.

The switching-field scaling was independently cross-validated using MuMax+ (Fig.~S19(e)).
Using a refined \(\SI{2}{nm}\) in-plane discretization, the independent
calculations give an approximate dependence
\(H_s\propto(c/b)^{0.79}\), compared with \(n=0.85\) for the MuMax3
phenomenological fit used here. The close agreement of the scaling exponents
supports the conclusion that the strong increase of \(H_s\) with
miniaturization is a robust geometry-dependent trend rather than a
solver-specific feature.

\subsection{Current-driven field-free switching}

The pulse-current switching simulations were performed using a heavy-metal/Py bilayer geometry. Unless otherwise stated, the heavy-metal layer was Ta with thickness \(t_{\mathrm{HM}}=\SI{5}{nm}\), placed underneath the Py free layer. Charge-current pulses were applied through the heavy-metal layer, and the SOT generated by the spin Hall effect was used to switch the Py magnetization. Thus, the \(I_s\) and \(J_s\) discussed here correspond to the current-driven Ta/Py bilayer device rather than a Py-only structure.

\begin{figure}[t!]
    \centering
    \includegraphics[width=\columnwidth, trim=0cm 10cm 0cm 0cm, clip]{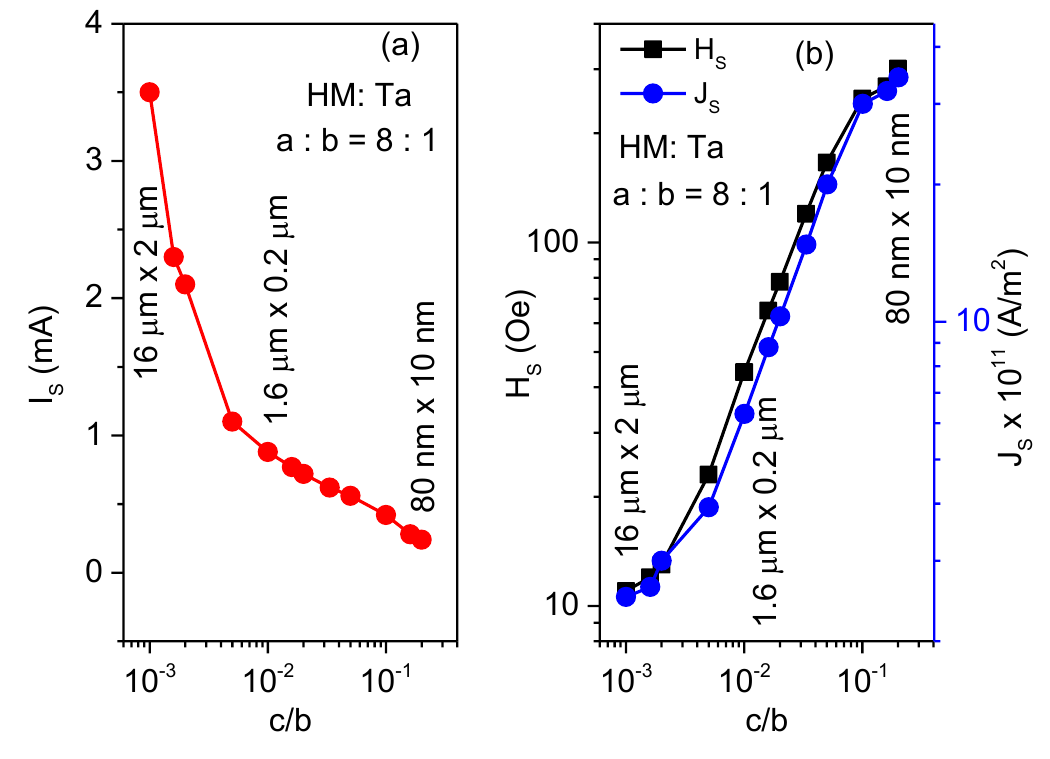}
    \caption{
    Current-driven size scaling for the crossed-ellipse device at fixed aspect ratio \(a:b=8:1\) using Ta as the heavy-metal layer. Panel (a) shows that the absolute switching current \(I_s\) decreases as the device is scaled down. Panel (b) shows that both the switching field \(H_s\) and the switching current density \(J_s\) increase with increasing \(c/b\). The vertical axis in panel (a) is plotted on a linear scale, whereas both vertical axes in panel (b) are plotted on logarithmic scales.
    }
    \label{fig:current_scaling}
    \vspace{-0.5cm}
\end{figure}

The current-driven scaling results are shown together with the switching-field scaling in Fig.~\ref{fig:current_scaling} and summarized in Table~\ref{tab:size_scaling}. At fixed aspect ratio \(a:b=8:1\), reducing the device size lowers the absolute \(I_s\), which is favorable for crossbar drivers. However, \(J_s\) increases with scaling and follows the same overall trend as \(H_s\). The same was also reproduced independently using MuMax+ confirming the same qualitative increase of \(J_s\) with \(c/b\) obtained in the MuMax3 calculations (Fig.~S19(f)). Device miniaturization therefore introduces a practical compromise: smaller cells require lower total current, which is favorable for write-driver scaling, but the constraint is transferred to larger current density, Joule heating, electromigration, and heavy-metal reliability concerns in SOT-based memory structures \cite{liuTa,garello,krizakovaReview,vlsiGarello}.

Similar current-driven switching simulations were also performed using Pt as the heavy-metal layer (Fig.~S1). In contrast to Ta, Pt has a positive spin Hall angle, \(\theta_{\mathrm{SH}}>0\), so the Oersted field \(B_{\mathrm{Oe}}\) assists the SOT-driven switching direction rather than opposing it. As a result, the Pt-based device shows a slightly reduced switching-current amplitude compared with the Ta/Py case. However, the qualitative scaling behavior remains unchanged: device miniaturization reduces \(I_s\), while increasing \(J_s\) and \(H_s\). This confirms that the observed scaling limitation is mainly controlled by geometry rather than by the choice of heavy metal, although the heavy-metal layer can still modify the absolute switching-current amplitude through the sign and efficiency of the spin Hall torque \cite{liuTa,paiAPL,manchonReview}.

\subsection{Aspect-ratio dependence at fixed major axis}

The second major part of the study focused on aspect-ratio tuning. In this simulation series, the major axis was fixed at \(a=\SI{1.6}{\micro m}\), while the minor axis \(b\) was varied. Therefore, increasing \(a/b\) corresponds to decreasing the width of the ellipse.

\begin{figure}[t!]
    \centering
    \includegraphics[width=\columnwidth, trim=0cm 0cm 1cm 0cm, clip]{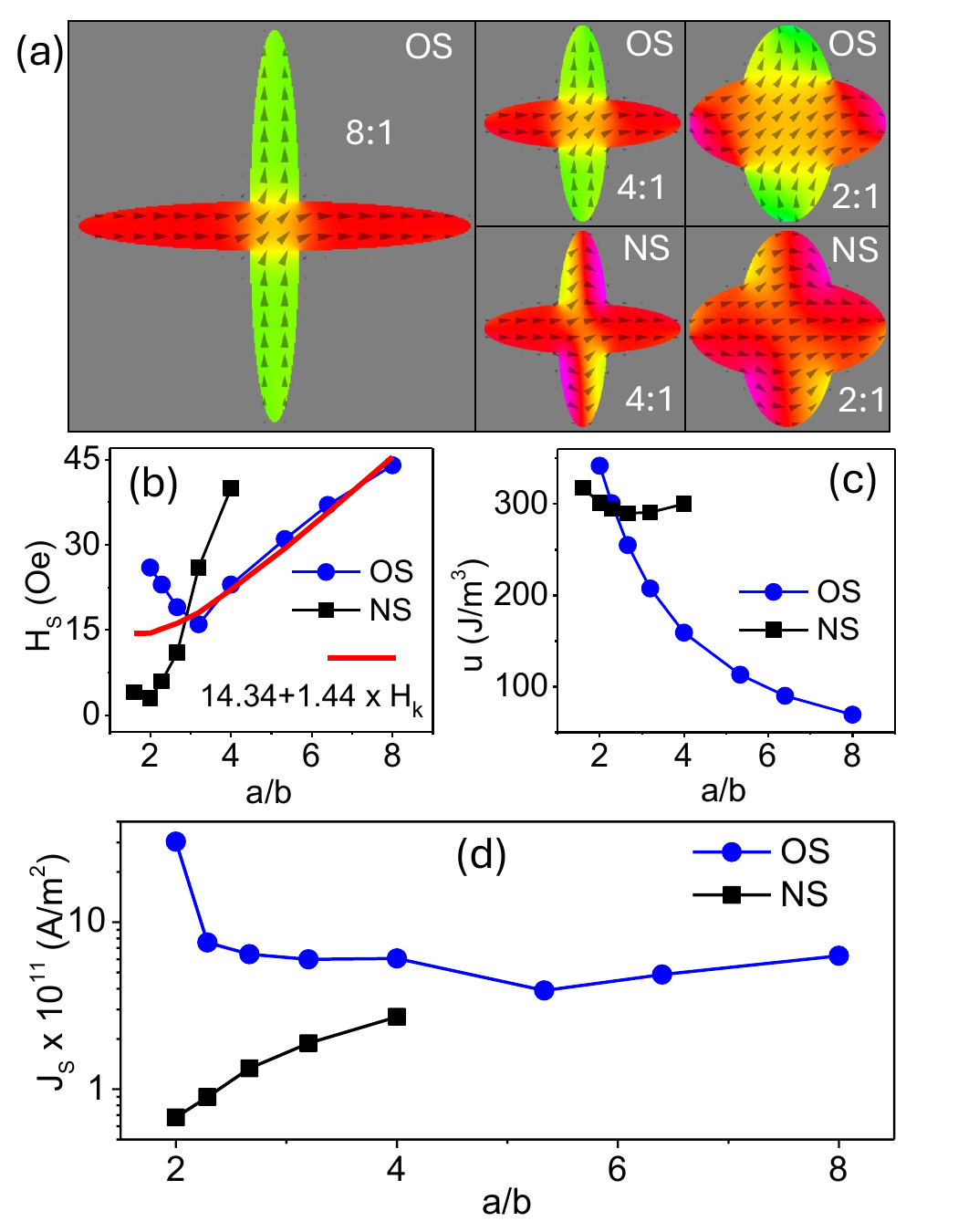}
    \caption{Aspect-ratio dependence of remanent states, switching field, remanent-state energy density, and SOT switching current density for crossed Py ellipses with fixed major axis \(a=\SI{1.6}{\micro m}\). Panel (a) compares ordinary states (OS) and newly stabilized states (NS) for selected aspect ratios. Panel (b) shows the nonmonotonic dependence of \(H_s\) on \(a/b\) for the OS branch and the lower \(H_s\) associated with the NS branch. The red curve shows the effective-anisotropy comparison \(H_s=14.34+1.44H_k\). Panel (c) shows that the remanent-state energy density of the OS branch decreases with increasing \(a/b\), while the NS branch remains nearly constant over the lower-aspect-ratio range. (d) Aspect-ratio dependence of the SOT switching current density \(J_s\) for the OS and NS branches, plotted in units of \(10^{11}~\mathrm{A/m^2}\).
    }
    \label{fig:aspect_ratio}
    \vspace{-0.5cm}
\end{figure}

The aspect-ratio-dependent \(H_k\), \(H_s\), and \(H'_s\) values are summarized in Table~\ref{tab:aspect_ratio}. The corresponding magnetic structures and aspect-ratio-dependent trends are shown in Fig.~\ref{fig:aspect_ratio}.

The ordinary-state (OS) switching field shows a clear nonmonotonic dependence on \(a/b\). Starting from \(a/b=8\), decreasing the aspect ratio reduces \(H_s\) from \(\sim\SI{44}{Oe}\) to a minimum of \(\sim\SI{16}{Oe}\) at \(a/b=3.2\). Further decrease in aspect ratio increases \(H_s\) again, reaching \(\sim\SI{26}{Oe}\) at \(a/b=2\). Thus, the lowest \(H_s\) for the original four-state switching path does not occur in the most elongated device, but rather at an intermediate aspect ratio.

For the high-aspect-ratio side of the OS branch, the switching field was compared with an effective crossed-ellipse anisotropy field \(H_k\), extracted from angular-response fitting using a Stoner-type two-crossed-ellipse model \cite{stonerwohlfarth,petersonIdzerdaSWFit,lahavPHEExtendedRange,morPHEShape}. The OS switching field follows the approximate relation \(H_s = H_0 + C H_k\), with \(H_0=\SI{14.34}{Oe}\), \(C=1.44\), and \(R^2=0.978\) over the range \(3.2\leq a/b\leq8\), as shown in Fig.~\ref{fig:aspect_ratio}(b). This indicates that the increasing \(H_s\) at large \(a/b\) is mainly controlled by the ordinary shape-anisotropy scale. At lower aspect ratios, however, the OS branch deviates from this linear trend and the NS branch becomes accessible. This deviation shows that the low-aspect-ratio reversal is no longer governed only by the ordinary anisotropy scale, but by a nonuniform multiminimum landscape created by the crossed-ellipse geometry. A representative angular-response fit used to extract \(H_k\) is shown in Fig.~S2.

\begin{table}[t!]
\centering
\renewcommand{\arraystretch}{1.2}
\setlength{\tabcolsep}{10pt}
\small
\caption{Switching field \((H_s)\), newly stabilized-state (NS) switching field \((H'_s)\), and effective crossed-ellipse anisotropy field \((H_k)\) for different aspect ratios at fixed major axis \(a=\SI{1.6}{\micro m}\). The \(H_k\) values were extracted from angular-response fitting using a Stoner-type crossed-ellipse model.}
\label{tab:aspect_ratio}
\begin{tabular}{ccccc}
\toprule
Device size & \(a/b\) & \(H_k\) & \(H_s\) & \(H'_s\) \\
 &  & (Oe) & (Oe) & (Oe) \\
\midrule
\(1.6\times0.2~\si{\micro m^2}\)  & 8.00 & 21.61 & 44.2 & -- \\
\(1.6\times0.25~\si{\micro m^2}\) & 6.40 & 14.82 & 37.3 & -- \\
\(1.6\times0.3~\si{\micro m^2}\)  & 5.33 & 10.44 & 30.8 & -- \\
\(1.6\times0.4~\si{\micro m^2}\)  & 4.00 & 5.40  & 23.1 & 39.8 \\
\(1.6\times0.5~\si{\micro m^2}\)  & 3.20 & 2.59  & 16.1 & 26.3 \\
\(1.6\times0.6~\si{\micro m^2}\)  & 2.67 & 1.29  & 19.3 & 11.1 \\
\(1.6\times0.7~\si{\micro m^2}\)  & 2.29 & 0.60  & 22.9 & 6.2 \\
\(1.6\times0.8~\si{\micro m^2}\)  & 2.00 & 0.08  & 26.1 & 3.1 \\
\(1.6\times1.0~\si{\micro m^2}\)  & 1.60 & 0.03  & --   & 3.8 \\
\bottomrule
\end{tabular}
\end{table}

The \(H_k\)-based comparison is applied only to the fixed-major-axis aspect-ratio series. In this case, the dominant change is the ordinary in-plane anisotropy scale of the crossed ellipse, and the OS switching field follows \(H_s=H_0+C H_k\) on the high-aspect-ratio branch. The same approach is not expected to describe the fixed-\(a/b\) size-scaling series, because \(a/b\) is held constant and the main change is instead the absolute magnetic length scale. In that regime, exchange, edge confinement, nonuniform reversal, and the increasing \(c/b\) ratio modify the switching pathway, so \(H_s\) is governed by finite-size micromagnetic effects rather than by the angular anisotropy scale alone.

At lower aspect ratios, additional remanent states become stabilized, giving rise to a separate newly stabilized-state (NS) switching branch denoted by \(H'_s\) in Table~\ref{tab:aspect_ratio}. This branch first appears at \(a/b=4.00\), where \(H'_s=\SI{39.8}{Oe}\), higher than the OS value of \(\SI{23.1}{Oe}\). At \(a/b=3.20\), \(H'_s\) decreases to \(\SI{26.3}{Oe}\), but still remains above the OS minimum. For still lower aspect ratios, the NS branch becomes more favorable for switching. For \(a/b=1.60\), the ordinary four-state branch was not observed under the present initialization and field-sweep protocol; instead, the device relaxed into the NS branch, with \(H'_s=\SI{3.8}{Oe}\). These results show that reducing the aspect ratio changes not only the magnitude of the switching field, but also the accessible remanent-state landscape and the preferred switching pathway.

The no-dead-layer calculation in Fig.~S3 preserves the same qualitative OS/NS branch structure. In this case, the OS branch follows the effective-anisotropy comparison \(H_s=\SI{20.75}{Oe}+1.07H_k\), where \(H_k\) is extracted from the same crossed-ellipse angular-response fitting used for the main-text aspect-ratio series. Thus, changing the active magnetic thickness mainly shifts the switching-field scale while preserving the branch structure and the interpretation that the high-aspect-ratio OS branch is governed primarily by the ordinary crossed-ellipse anisotropy scale.

The remanent-state energy density, shown in Fig.~\ref{fig:aspect_ratio}(c), decreases monotonically with increasing \(a/b\) for the OS branch, consistent with stronger long-axis alignment driven by shape anisotropy \cite{osborn,aharoni,cowburn}. In contrast, the NS branch has a nearly constant energy density over the range where it appears, suggesting a distinct remanent-state family with weaker sensitivity to small changes in \(a/b\).

The aspect-ratio dependence of the SOT \(J_s\) is shown in Fig.~\ref{fig:aspect_ratio}(d). The OS branch requires a larger \(J_s\) overall and shows a broad minimum in the intermediate aspect-ratio range. The NS branch requires a much lower \(J_s\) in the low-aspect-ratio range where these states exist, although \(J_s\) increases as \(a/b\) increases. These results show that the optimum geometry depends on the targeted remanent-state branch: the original four-state pathway is optimized near \(a/b=3.2\) for the \(a=\SI{1.6}{\micro m}\) geometry, whereas NS states are more easily accessed at lower aspect ratios.

Supplementary fixed-major-axis calculations show that the same qualitative OS/NS branch structure persists for \(a=\SI{3.2}{\micro m}\) and \(a=\SI{0.8}{\micro m}\), but that the switching-field scale, the location of the OS minimum, and the effective-anisotropy relation \(H_s=H_0+C H_k\) depend on absolute size (Figs.~S4 and S5). Thus, the \(a/b=3.2\) optimum is specific to the \(a=\SI{1.6}{\micro m}\) geometry rather than a universal optimum for all crossed ellipses. The same size dependence appears in the angle-dependent state-count phase map in Fig.~S9, where only field-off configurations that persist over at least two consecutive \(\SI{5}{^\circ}\)-spaced field angles are counted as robust plateaus. Under this criterion, high-aspect-ratio geometries return to the ordinary four-state landscape, whereas lower-aspect-ratio geometries support additional robust remanent plateaus.

\subsection{Angle-dependent twelve-state response}

\begin{figure*}[t!]
    \centering
    \includegraphics[width=1.4\columnwidth]{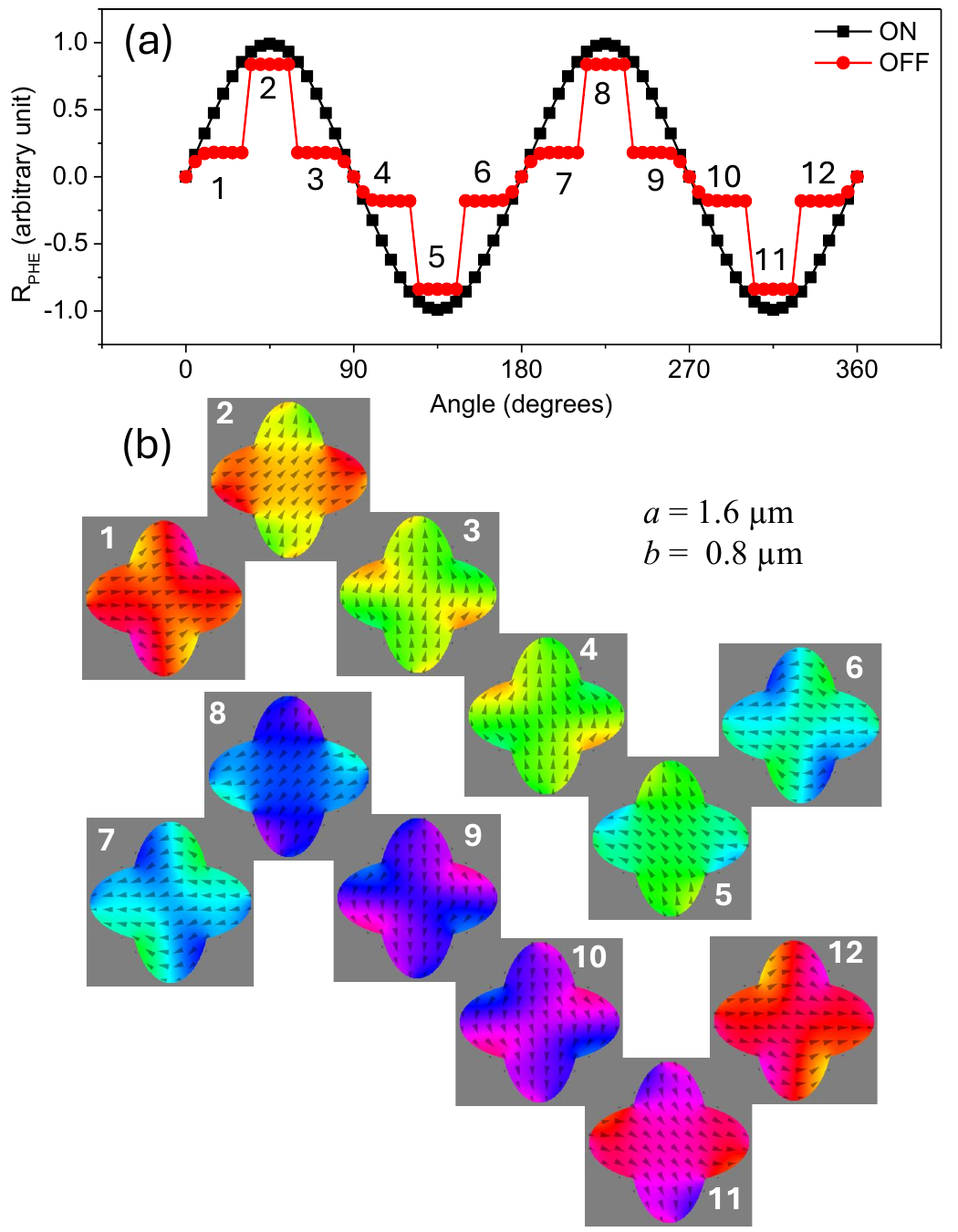}
    \caption{
    (a) Angle-dependent simulated planar Hall response of the \(1.6~\si{\micro m}\times0.8~\si{\micro m}\) crossed-ellipse device measured at an applied field of \(\SI{100}{Oe}\). The planar Hall response was calculated as \(R_{\mathrm{PHE}}=2\langle m_x\rangle\langle m_y\rangle\). The field-on response, \(R_{\mathrm{PHE,ON}}\), was obtained while the magnetic field was applied at each in-plane angle, whereas the field-off response, \(R_{\mathrm{PHE,OFF}}\), was obtained after removing the field and relaxing the device into its remanent state. The twelve labeled plateaus indicate twelve accessible field-off remanent configurations in the lower-aspect-ratio geometry, as shown in panel (b).
    }
    \label{fig:twelve_state_response}
\end{figure*}

As anticipated from the low-aspect-ratio NS branch in Section~3.3, these geometries can host more than the ordinary four remanent states. To resolve the accessible state landscape explicitly, the angle-dependent field-on and field-off response was calculated for the \(1.6~\si{\micro m}\times0.8~\si{\micro m}\) device, as shown in Fig.~\ref{fig:twelve_state_response}. The simulated planar Hall response was calculated from the spatially averaged in-plane magnetization components as \(R_{\mathrm{PHE}}=2\langle m_x\rangle\langle m_y\rangle\), where \(\langle m_x\rangle\) and \(\langle m_y\rangle\) are the spatial averages of the normalized magnetization components. The simulations were performed using an applied field of \(\SI{100}{Oe}\). For each in-plane field angle, the magnetization was first relaxed under the applied field to obtain the field-on state. The field was then removed, and the device was relaxed again to obtain the corresponding field-off remanent state.

The field-on response follows the applied field direction and shows an approximately continuous angular dependence. In contrast, the field-off response separates into discrete plateaus after relaxation. The plateau analysis identifies twelve remanent states, with approximate angular positions \(6.0^\circ\), \(45.0^\circ\), \(84.1^\circ\), \(95.9^\circ\), \(135.0^\circ\), \(174.0^\circ\), \(185.9^\circ\), \(225.0^\circ\), \(264.0^\circ\), \(275.9^\circ\), \(315.0^\circ\), and \(354.0^\circ\), as summarized in Table~\ref{tab:twelve_state_energy}. The angular separations are not uniform; most neighboring states are separated by approximately \(39^\circ\), while four neighboring pairs are separated by only approximately \(12^\circ\). This nonuniform spacing indicates that the states are not generated by simple macrospin rotation, but by a geometry-dependent energy landscape involving the two crossed arms and the central overlap region. Both the refined MuMax3 calculation and the independent MuMax+ calculation with \(\SI{2}{nm}\) cell size reproduce the same multistate topology despite differences in solver implementation (Fig.~S19(g)). This agreement provides strong evidence that the additional remanent configurations arise from the crossed-ellipse energy landscape rather than from a particular discretization or micromagnetic solver.

The corresponding field-off energy density was also extracted for each remanent plateau. The twelve states form two energy-density families. The eight canted states located close to the principal axes have nearly identical energy density, \(E_{\mathrm{tot}}\approx\SI{301.30}{J/m^3}\), whereas the four diagonal states near \(45^\circ\), \(135^\circ\), \(225^\circ\), and \(315^\circ\) have a higher energy density, \(E_{\mathrm{tot}}\approx\SI{341.55}{J/m^3}\). The energy-density separation between these two families is approximately \(\SI{40.25}{J/m^3}\). This confirms that the field-off plateaus correspond to relaxed remanent configurations under the zero-field protocol, although the energy of a minimum is not equivalent to the thermal activation barrier separating neighboring states. It is important to distinguish the number of remanent configurations from the number of distinct energy values. Symmetry-related magnetization
configurations can have the same total energy while remaining distinct field-off states. Thus, the twelve observed plateaus correspond to twelve accessible micromagnetic configurations, but they need not correspond to twelve nondegenerate energies or twelve thermally independent memory basins.

\begin{table*}[t!]
\centering
\renewcommand{\arraystretch}{1.2}
\setlength{\tabcolsep}{10pt}
\small
\caption{Field-off remanent-state parameters and estimated MTJ resistance levels for the \(1.6~\si{\micro m}\times0.8~\si{\micro m}\) crossed-ellipse device. The plateau angle \(\phi_{\mathrm{OFF}}\) was calculated from the spatially averaged magnetization components. The MTJ resistance estimate assumes an in-plane reference layer at \(\phi_{\mathrm{ref}}=35.4^\circ\), using \(q_i=\langle m_x\rangle_i\cos\phi_{\mathrm{ref}}+\langle m_y\rangle_i\sin\phi_{\mathrm{ref}}\) and \(R_i/R_P=(1+P)/(1+Pq_i)\), with \(P=\mathrm{TMR}/(\mathrm{TMR}+2)\).}
\label{tab:twelve_state_energy}
\begin{tabular}{cccccccc}
\toprule
State & Range & \(\phi_{\mathrm{OFF}}\) & \(R_{\mathrm{PHE}}\) & \(E_{\mathrm{tot}}\) & \(q_i\) & \multicolumn{2}{c}{\(R_i/R_P\)} \\
 & \((^\circ)\) & \((^\circ)\) &  & \((\mathrm{J\,m^{-3}})\) & & \(100\%\) TMR & \(200\%\) TMR \\
\midrule
S1  & 10--30   & 6.0   & 0.180  & 301.30 &  0.4657 & 1.1542 & 1.2167 \\
S2  & 35--55   & 45.0  & 0.838  & 341.55 &  0.5186 & 1.1368 & 1.1911 \\
S3  & 60--80   & 84.1  & 0.179  & 301.29 &  0.3510 & 1.1937 & 1.2761 \\
S4  & 100--120 & 95.9  & -0.179 & 301.29 &  0.2653 & 1.2250 & 1.3243 \\
S5  & 125--145 & 135.0 & -0.838 & 341.55 & -0.0928 & 1.3759 & 1.5730 \\
S6  & 150--170 & 174.0 & -0.179 & 301.30 & -0.4005 & 1.5388 & 1.8756 \\
S7  & 190--210 & 185.9 & 0.179  & 301.30 & -0.4656 & 1.5783 & 1.9552 \\
S8  & 215--235 & 225.0 & 0.838  & 341.55 & -0.5186 & 1.6120 & 2.0251 \\
S9  & 240--260 & 264.0 & 0.180  & 301.29 & -0.3511 & 1.5101 & 1.8194 \\
S10 & 280--300 & 275.9 & -0.179 & 301.29 & -0.2652 & 1.4626 & 1.7293 \\
S11 & 305--325 & 315.0 & -0.838 & 341.55 &  0.0929 & 1.2933 & 1.4334 \\
S12 & 330--350 & 354.0 & -0.180 & 301.30 &  0.4005 & 1.1763 & 1.2497 \\
\bottomrule
\end{tabular}
\end{table*}

To estimate the corresponding MTJ resistance levels, the twelve remanent states were treated as the free-layer states of an in-plane-anisotropy MTJ. The estimate follows the standard spin-dependent tunneling picture, in which the conductance depends on the relative orientation of the two magnetic electrodes \cite{julliere,slonczewskiMTJ,moodera}. For a reference layer oriented at \(\phi_{\mathrm{ref}}\), the effective magnetic projection was calculated as \(q_i=\langle m_x\rangle_i\cos\phi_{\mathrm{ref}}+\langle m_y\rangle_i\sin\phi_{\mathrm{ref}}\). The MTJ resistance was then estimated using the conductance-based angular dependence \(R_i/R_P=(1+P)/(1+Pq_i)\), where \(P=\mathrm{TMR}/(\mathrm{TMR}+2)\). For the illustrative case of \(\phi_{\mathrm{ref}}=35.4^\circ\), the twelve remanent states give twelve distinct projected resistance levels for both \(\mathrm{TMR}=100\%\) and \(\mathrm{TMR}=200\%\), as listed in Table~\ref{tab:twelve_state_energy}. The reference-layer angle was chosen from a scan of the minimum adjacent projected resistance-level spacing; this scan gives equivalent maxima near \(35.4^\circ\) and \(54.6^\circ\), reflecting the approximate symmetry of the twelve-state landscape (Fig.~S6). This shows that the twelve accessible field-off configurations can, in principle, be converted into multiple projected MTJ resistance levels if the reference-layer direction is chosen away from high-symmetry axes. However, the projected resistance-level count should be distinguished from the thermal memory-level count. As shown by the representative string-method calculations in the Supplementary Information and discussed below, the closest canted pairs are weakly separated and may merge thermally, reducing the number of robust effective basins for the present geometry.

For \(\mathrm{TMR}=100\%\) and \(\phi_{\mathrm{ref}}=35.4^\circ\), the calculated levels span \(R_i/R_P=1.1368\text{--}1.6120\). The minimum adjacent level spacing is \(\Delta(R/R_P)_{\min}\approx0.0174\), corresponding to approximately \(1.5\%\) of the lower neighboring level. Increasing the assumed TMR to \(200\%\) increases the calculated span to \(R_i/R_P=1.1911\text{--}2.0251\) and the minimum adjacent spacing to \(\Delta(R/R_P)_{\min}\approx0.0256\). Thus, resolving all twelve levels would require resistance noise and device-to-device variation to remain well below the relevant spacing scale. This estimate emphasizes that the twelve magnetic states define the available remanent-state landscape, whereas the practically usable number of MTJ readout levels will depend on the resistance margin of the fabricated junction.

The low-aspect-ratio state-count comparison in the Supplementary Information shows that this twelve-state response is not unique to the \(1.6~\si{\micro m}\times0.8~\si{\micro m}\) device. The larger \(3.2~\si{\micro m}\times1.6~\si{\micro m}\) device also supports twelve well-resolved field-off states, whereas the smaller \(0.8~\si{\micro m}\times0.4~\si{\micro m}\) device supports eight states (Figs.~S7 and S8). The broader phase map in Fig.~S9 further confirms that the number of resolved states is controlled by both aspect ratio and absolute major-axis size.
\vspace{-0.25cm}
\section{Discussion}
\vspace{-0.25cm}
The simulations identify the main geometry constraints for using crossed-ellipse Py structures as candidate multistate MTJ cells. Size reduction and aspect-ratio tuning affect different design metrics. Scaling the device at fixed \(a:b=8:1\) lowers the absolute switching current, which is favorable for dense arrays and write-current delivery, but it also increases both \(H_s\) and \(J_s\). Thus, miniaturization improves cell density only while the resulting field and current-density requirements remain compatible with reliable operation \cite{ross,krizakovaReview}.

The aspect-ratio results show that the design optimum is branch- and size-dependent. For the \(a=\SI{1.6}{\micro m}\) series, the OS branch gives a low-field four-state operating point near intermediate aspect ratio, while lower aspect ratios favor the NS branch and increase the number of accessible remanent states. The supplementary low-aspect-ratio comparison and state-count phase map in Figs.~S7--S9 extend this conclusion by showing that the transition from four-state to multistate behavior depends on absolute major-axis size as well as \(a/b\). This means that crossed-ellipse cells cannot be scaled by preserving aspect ratio alone; the lateral size, selected state branch, and write-current-density limit must be optimized together.

A useful way to view the crossed ellipse is as two shape-anisotropic arms coupled through a shared overlap region. At high aspect ratio, each arm is strongly confined along its long axis, and the overlap region mainly mediates the four diagonal remanent states. Reducing the aspect ratio weakens the arm anisotropy, allowing the arm magnetizations and the overlap region to cant relative to one another. This produces additional local minima near the principal axes, as observed in the \(1.6~\si{\micro m}\times0.8~\si{\micro m}\) geometry. The nonuniform angular spacing and the two energy-density families in Table~\ref{tab:twelve_state_energy} therefore indicate a discretized energy landscape shaped by arm magnetization, overlap-region canting, and magnetostatic coupling, rather than continuous macrospin rotation. If the aspect ratio is reduced further or the absolute device size is decreased, exchange and overlap-region coupling can merge some of the nearby canted minima, reducing the number of thermally distinguishable basins.

The projection-based MTJ estimate suggests that this magnetic state landscape can be converted into multiple electrical resistance levels, but the estimate should be viewed as an upper-bound readout scenario. For the chosen \(\phi_{\mathrm{ref}}=35.4^\circ\), the twelve remanent states give distinct calculated resistance levels for both \(\mathrm{TMR}=100\%\) and \(\mathrm{TMR}=200\%\). The supplementary resistance-level comparison in Fig.~S10 extends this estimate to the other \(a/b=2\) devices and shows that the \(3.2~\si{\micro m}\times1.6~\si{\micro m}\) and \(1.6~\si{\micro m}\times0.8~\si{\micro m}\) devices provide twelve projected levels, whereas the \(0.8~\si{\micro m}\times0.4~\si{\micro m}\) device provides eight. However, several levels are closely spaced, so in a real MTJ the usable number of levels will depend on the TMR ratio, readout noise, reference-layer orientation, thermal fluctuations, and device-to-device variation. Therefore, the present calculation identifies a promising multistate free-layer landscape and provides a first projected level-spacing estimate, while a full device-level resistance-margin analysis remains a future requirement.

The Py simulations should be viewed as a geometry-design map rather than as a material-specific prescription. In practical MTJ stacks, CoFeB/MgO/CoFeB stacks are especially attractive because crystalline MgO barriers enable coherent tunneling and experimentally demonstrated large room-temperature TMR ratios, ranging from above \(100\%\) in scaled perpendicular-anisotropy MTJs to several hundred percent in optimized in-plane CoFeB/MgO/CoFeB junctions \cite{parkinMgOGiantTMR,hayakawaCoFeBMgO260,ikedaCoFeBMgO604,ikedaPMACoFeBMgO}. Such larger TMR ratios would improve the electrical separation between projected resistance levels. The supplementary CoFeB checks show that the OS/NS branch structure and the low-aspect-ratio twelve-state response persist when CoFeB-like magnetic parameters are used, although the switching-field scale changes (Figs.~S12 and S13). This is expected because the magnetostatic and shape-anisotropy energy scales depend on \(M_s\), while thermal stability is controlled by the relevant energy barriers \(E_b\), which may also change with magnetic volume, anisotropy, and stack details. Therefore, moving from Py to CoFeB may improve MTJ readout and retention margins, but it also shifts the switching-field and current-density requirements. A full CoFeB/MgO optimization should therefore include the actual stack parameters, interfacial anisotropy, damping, spin Hall efficiency, and reference-layer coupling.

Thermal stability is the next critical requirement for using the accessible
remanent configurations as nonvolatile memory levels. The energy-density
values of the relaxed states do not determine the saddle-point barriers
between them, and the relevant retention scale is set by the lowest available
escape pathway. Representative zero-field string-method calculations using
the full micromagnetic configurations confirm that the closest canted states
are only weakly separated (Fig.~S11(a)). In the MuMax3 calculations, the adjacent S3--S4
path relaxes to an essentially barrierless connection, whereas selected
diagonal-to-canted and diagonal-to-diagonal endpoint calculations retain
substantially larger apparent barriers.

Independent MuMax+ minimum-energy-path calculations support the weak thermal
separation of the closest canted states, but also reveal an important
limitation of direct endpoint-to-endpoint barrier estimates (Figs.~S11(b-d)). A string
connecting two non-neighbouring minima does not necessarily identify the
lowest physical escape pathway if one or more metastable configurations lie
between the selected endpoints. In the independently calculated landscape,
for example, a nominal direct transition between diagonal configurations can
proceed through intermediate canted states as a sequence of lower-barrier
steps. The retention barrier of a given remanent state must therefore be
identified from its lowest neighbouring-state escape pathway rather than from
a direct path connecting distant endpoints.

Accordingly, the twelve field-off plateaus should be interpreted as twelve
accessible remanent configurations rather than as twelve nonvolatile memory
states. The present minimum-energy-path calculations establish a strongly
nonuniform hierarchy of inter-state barriers and show that some neighbouring
configurations are thermally connected. However, they do not justify assigning
a unique eight-state nonvolatile count to the present geometry without first
constructing the complete network of neighbouring-state transition barriers.
The \(1.6~\si{\micro m}\times0.8~\si{\micro m}\) device should therefore be
regarded as a demonstration of a geometry-generated multistate landscape
whose room-temperature retention requires further optimization of aspect
ratio, magnetic thickness, anisotropy, material parameters, and stack design.
Thus, the number of practically usable MTJ resistance levels will be set by the lowest inter-state barriers, not only by the number of field-off plateaus or projected resistance levels.

The connection to earlier experiments remains important. The four-state MTJ demonstrated in Ref.~\cite{dasAPL} provides an experimental anchor for the crossed-ellipse concept, showing distinct resistance states and field-free SOT switching in a Ta-based structure. In that experiment, the crossed-ellipse free layer had principal axes of \(8~\si{\micro m}\) and \(1~\si{\micro m}\), and a field of \(\SI{7}{Oe}\) applied and removed at different in-plane angles was sufficient to switch the free layer between its four remanent states. For the same nominal \(8:1\) crossed-ellipse geometry, the present isolated-free-layer simulation gives \(H_s=\SI{13.1}{Oe}\). The simulation therefore overestimates the experimental value by approximately a factor of two, which is plausible for an idealized free-layer model compared with a full MTJ stack that includes the top single-ellipse layer, interlayer coupling, stray field, fabrication-induced edge rounding, and material variations. In particular, stray-field and orange-peel-coupling contributions in the experimental MTJ can bias the crossed-ellipse free-layer energy landscape and shift the apparent switching field \cite{dasAPL}. Therefore, the simulations should be interpreted as geometry-dependent design trends rather than absolute predictions of the measured switching field.

Reference-layer stray field and coupling can also modify the apparent multistate response if the reference layer produces an uncompensated in-plane bias field. In a practical MTJ design, this perturbation can be reduced by using a small single-ellipse reference layer centered on the crossed-ellipse free layer, so that the reference layer samples the central overlap region while minimizing edge-generated stray fields over the extended arms~\cite{dasAPL}. The stray field can be further suppressed by using a synthetic-antiferromagnetic (SAF) reference stack, for which the dipolar fields from the coupled magnetic layers largely compensate; SAF pinned/reference layers are commonly used in MTJs to reduce reference-layer stray field and free-layer offset fields~\cite{bandieraSAFReference,ikedaPerpendicularMTJ,saitoIrReSAF}. As a conservative sensitivity check, supplementary angular sweeps were performed with static in-plane bias fields applied along the optimized reference-layer direction, \(\phi_{\mathrm{ref}}=35.4^\circ\). A weak \(\SI{2}{Oe}\) bias preserves the main multistate response while beginning to merge the closest NS canted pairs. At \(\SI{5}{Oe}\), several neighboring configurations merge and the accessible response is reduced to six field-off states, while at \(\SI{10}{Oe}\) even the ordinary-state plateaus are strongly affected. These results show that the low-aspect-ratio multistate landscape is not solely an artifact of exact fourfold symmetry, but they also indicate that uncompensated reference-layer bias should be minimized through stack and reference-layer design.

For crossbar implementation, the single-cell design must also be evaluated in an array environment. A preliminary 3\(\times\)3 array-coupling check in the Supplementary Information indicates that nearest-neighbor magnetostatic coupling is weak for the high-aspect-ratio ordinary-state geometry at the tested pitch \(p/a=1.25\): the center-cell switching field changes from \(\SI{42.2}{Oe}\) to \(\SI{42.6}{Oe}\). The lower-aspect-ratio NS branch is more sensitive, changing from \(\SI{2.9}{Oe}\) in the isolated-cell calculation to \(\SI{4.5}{Oe}\) at \(p/a=1.25\), but this perturbation decreases to \(\SI{3.6}{Oe}\) when the pitch is increased to \(p/a=2.5\). Thus, the single-cell trends appear qualitatively robust against the tested nearest-neighbor magnetostatic environment, although full crossbar use will require additional optimization of pitch, thermal stability, write selectivity, current distribution, and readout margin.

Experimentally, the quantitative optimum may shift with fabrication conditions. Lithographic edge roughness can become important as the device width approaches the sub-100-nm scale, while the effective magnetic thickness can vary with interface quality and the assumed dead-layer correction. The no-dead-layer comparison in Fig.~S3 shows that changing the effective magnetic thickness mainly shifts the switching-field scale while preserving the main aspect-ratio trend.
In addition, the lithographic-roughness check in the Supplementary Information shows that a \(\SI{5}{nm}\) rough-edge geometry changes \(H_s\) by only \(4\text{--}6\%\) for the tested \(a=\SI{1.6}{\micro m}\) Py devices with \(a/b=8,6,\) and \(4\), without changing the qualitative aspect-ratio trend. Therefore, the simulated results should be viewed as geometry-design guidelines rather than fixed universal dimensions, but the main trends are not dominated by small edge-geometry perturbations at the level tested here.

\vspace{-0.25cm}
\section{Conclusion}
\vspace{-0.25cm}
Micromagnetic simulations show that crossed-ellipse Py free layers are governed by coupled size and aspect-ratio effects. At fixed \(a:b=8:1\), scaling from \(16~\si{\micro m}\times2~\si{\micro m}\) to \(80~\si{nm}\times10~\si{nm}\) lowers the absolute switching current, but increases both \(H_s\) and \(J_s\), defining the practical write-cost of miniaturization.

At fixed major axis \(a=\SI{1.6}{\micro m}\), the ordinary four-state branch has its lowest switching field at an intermediate aspect ratio. Lower-aspect-ratio devices stabilize additional remanent states with lower switching fields and lower SOT current-density requirements. The effective-anisotropy comparison shows that the high-aspect-ratio OS branch is governed mainly by the ordinary crossed-ellipse anisotropy scale, while deviations at low aspect ratio mark the onset of a nonuniform multistate landscape. The \(1.6~\si{\micro m}\times0.8~\si{\micro m}\) device supports twelve field-off remanent plateaus, and supplementary state-count mapping shows that the resolved state count depends on both aspect ratio and absolute size.

Projection-based MTJ estimates show that these configurations can, in principle, generate multiple electrical readout levels for an optimized reference-layer orientation. Minimum-energy-path calculations, however, show that the accessible plateau count is larger than
the thermally independent memory-state count: the closest canted
configurations are separated by very small barriers, and multistep pathways
through intermediate minima can reduce the escape barrier of nominally more
distant states. The present low-aspect-ratio geometry should therefore be
viewed as a demonstrated multistate energy landscape rather than as an already
optimized nonvolatile multilevel memory cell. Further optimization of
geometry, magnetic thickness, material parameters, and reference-stack design
is required to combine the enhanced state count with room-temperature
retention.
CoFeB checks indicate that the OS/NS branch structure and low-aspect-ratio multistate response are geometry-driven and persist for MTJ-relevant material parameters. Together, these results identify crossed-ellipse free layers as candidate multistate elements for spintronic crossbars, with robust four-state operation favored at intermediate aspect ratio and beyond-four-state operation favored at lower aspect ratio.
\vspace{-0.25cm}
\section*{Supplementary Information}
\vspace{-0.25cm}
Supplementary Information is available from the journal's online library or from the authors. It includes current-driven size scaling using Pt as the heavy-metal layer (Fig.~S1), effective crossed-ellipse anisotropy-field fitting and aspect-ratio-dependent switching without the magnetic dead-layer correction (Figs.~S2--S3), aspect-ratio-dependent switching for larger and smaller fixed major axes (Figs.~S4--S5), reference-layer-angle optimization for projected MTJ readout (Fig.~S6), low-aspect-ratio state-count and estimated MTJ resistance levels for \(a/b=2\) devices with different absolute sizes (Figs.~S7--S8), a state-count phase map (Fig.~S9), estimated MTJ resistance-level spacing (Fig.~S10), representative string-method and independent MuMax+ minimum-energy-path calculations of inter-state barriers (Fig.~S11), CoFeB parameter checks (Figs.~S12--S13), preliminary array-coupling checks (Fig.~S14), bias-field checks for reference-layer-induced asymmetry (Figs.~S15--S17), lithographic edge-roughness sensitivity (Fig.~S18), and independent MuMax+ cross-validation of mesh convergence, switching-field scaling, switching-current-density scaling, and the low-aspect-ratio twelve-state response (Fig.~S19).
\vspace{-0.25cm}
\section*{Acknowledgments}
\vspace{-0.25cm}
L.K. and F.A.A. acknowledge support from the Multispin.AI project, which has received funding from EU under grant agreement number 101130046. ``Funded by the European Union. Views and opinions expressed are however those of the author(s) only and do not necessarily reflect those of the European Union or the European Innovation Council and SMEs Executive Agency (EISMEA). Neither the European Union nor the granting authority can be held responsible for them.''
The authors thank Konstantin Zvezdin for his critical reading of the manuscript and for valuable suggestions.
\vspace{-0.25cm}
\section*{Author Contributions}
\vspace{-0.25cm}
All authors contributed to the conceptualization of the project. A.G. performed the MuMax3 simulations and data analysis, and wrote the original manuscript under the supervision of L.K. A.Z. contributed to the physical interpretation, conceptual development, and implementation of the study. C.D. and T.C. independently revalidated the principal results using MuMax+ under the supervision of F.A.A. A.G. and L.K. revised and edited the manuscript, with additional critical input from the other authors. All authors reviewed and approved the final version of the manuscript.
\vspace{-0.25cm}
\section*{Data availability statement}
\vspace{-0.25cm}

The processed data supporting the figures and tables in this work, together
with representative MuMax3 input scripts, are openly available in Zenodo
\cite{GhoshZenodo2026}.
The corresponding GitLab repository is available at
\href{https://gitlab.com/arupghoshphy/simulation-codes-and-processed-data-for-crossed-ellipse-multistate-mtj-free-layers}
{https://gitlab.com/arupghoshphy/simulation-codes-and-processed-data-for-crossed-ellipse-multistate-mtj-free-layers}
and will provide updated simulation files, analysis scripts, and additional
supporting resources as needed. Additional simulation and analysis files are
also available from the corresponding author upon reasonable request.


\bibliography{references}

\end{document}


\maketitle

\tableofcontents

\vspace{1cm}
\newpage
\section{Current-driven size scaling using Pt as the heavy-metal layer}

\begin{figure}[H]
    \centering
    \includegraphics[width=0.5\textwidth, trim=0cm 9.5cm 0cm 0cm, clip]{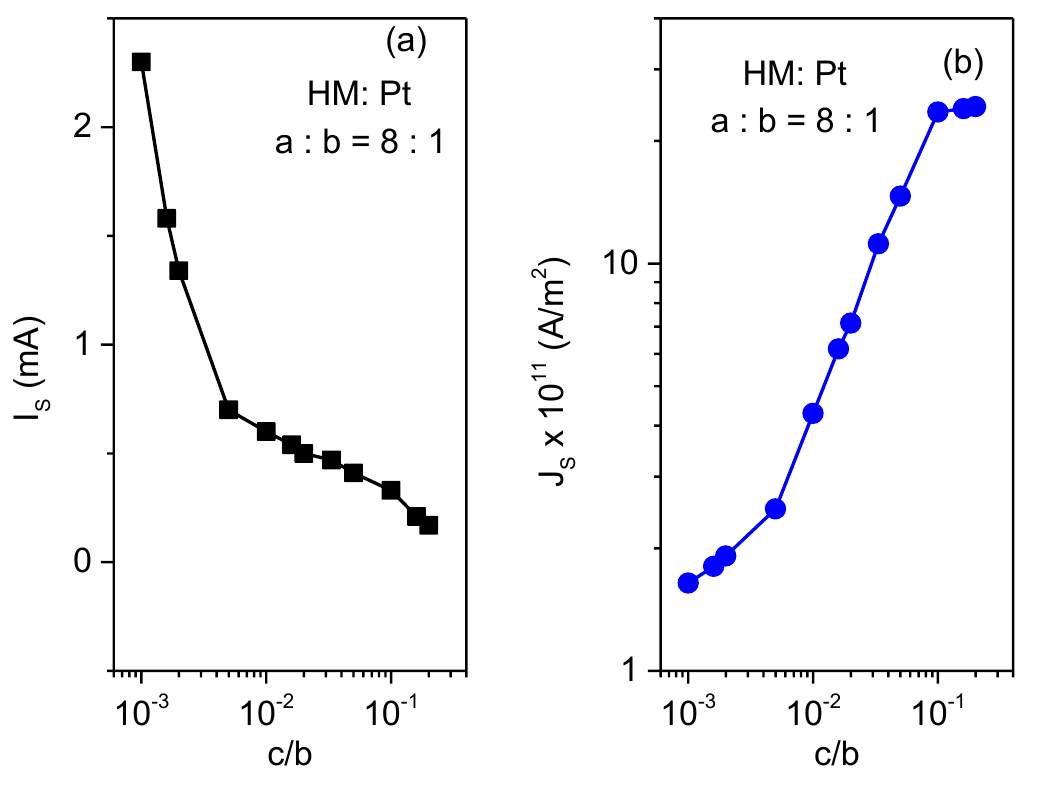}
    \caption{
    Current-driven size scaling using Pt as the heavy-metal layer at fixed aspect ratio \(a:b=8:1\). Panel (a) shows the absolute switching current \(I_s\) as a function of \(c/b\). Panel (b) shows the corresponding SOT switching current density \(J_s\).
    }
    \label{fig:pt_scaling}
\end{figure}

Figure~\ref{fig:pt_scaling} shows the current-driven size scaling for the crossed-ellipse device when Pt is used as the heavy-metal layer. The aspect ratio was kept fixed at \(a:b=8:1\), and the normalized thickness-to-width ratio \(c/b\) was increased by reducing the lateral dimensions while keeping the magnetic layer thickness fixed. The absolute switching current \(I_s\) decreases as the device is scaled down, whereas the corresponding SOT switching current density \(J_s\) increases. This behavior is qualitatively similar to the Ta-based case discussed in the main text, showing that the size-scaling trend is mainly controlled by geometry. The Pt case gives a slightly lower current amplitude because the sign of the Pt spin Hall angle makes the Oersted-field contribution assist the effective SOT-driven switching direction.

\section{Effective crossed-ellipse anisotropy-field comparison}

To provide a physical reference for the aspect-ratio dependence of the ordinary-state (OS) branch, the switching fields were compared with an effective crossed-ellipse anisotropy field \(H_k\). The fitting procedure is based on a Stoner--Wohlfarth-type angular-energy model, in which the magnetization direction is obtained by minimizing the sum of an anisotropy term and a Zeeman term for each applied-field angle \cite{stonerwohlfarth,petersonIdzerdaSWFit}. Similar angular-energy approaches are commonly used to describe planar Hall responses of shape-anisotropic magnetic elements, where the transverse signal follows the in-plane magnetization angle \cite{lahavPHEExtendedRange,morPHEShape}.

For a single ellipse, the dominant shape anisotropy can be represented by a twofold angular term. For the crossed-ellipse geometry, the two orthogonal arms produce an effective crossed-ellipse angular landscape. In the simplified fitting model used here, the field-dependent magnetic angle \(\phi\) was obtained by minimizing an effective energy of the form
\[
u(\phi)=
-\mu_0 M_s H\cos(\phi-\theta_H)
+\frac{1}{2}\mu_0 M_s H_k f_{\mathrm{CE}}(\phi),
\]
where \(H\) and \(\theta_H\) are the magnitude and in-plane angle of the applied field, \(M_s\) is the saturation magnetization, and \(f_{\mathrm{CE}}(\phi)\) is the crossed-ellipse angular anisotropy function used to represent the two orthogonal arms. The planar Hall response was then calculated from the fitted magnetization direction as
\[
R_{\mathrm{PHE}}\propto \sin 2\phi .
\]
The value of \(H_k\) was adjusted until the calculated angular response reproduced the simulated field-on planar Hall curve. Therefore, \(H_k\) should be interpreted as an effective ordinary anisotropy field of the crossed-ellipse structure, not as a direct switching field or a full micromagnetic energy barrier.

\begin{figure}[H]
    \centering
    \includegraphics[width=0.5\textwidth, trim=0cm 9.2cm 0cm 0cm, clip]{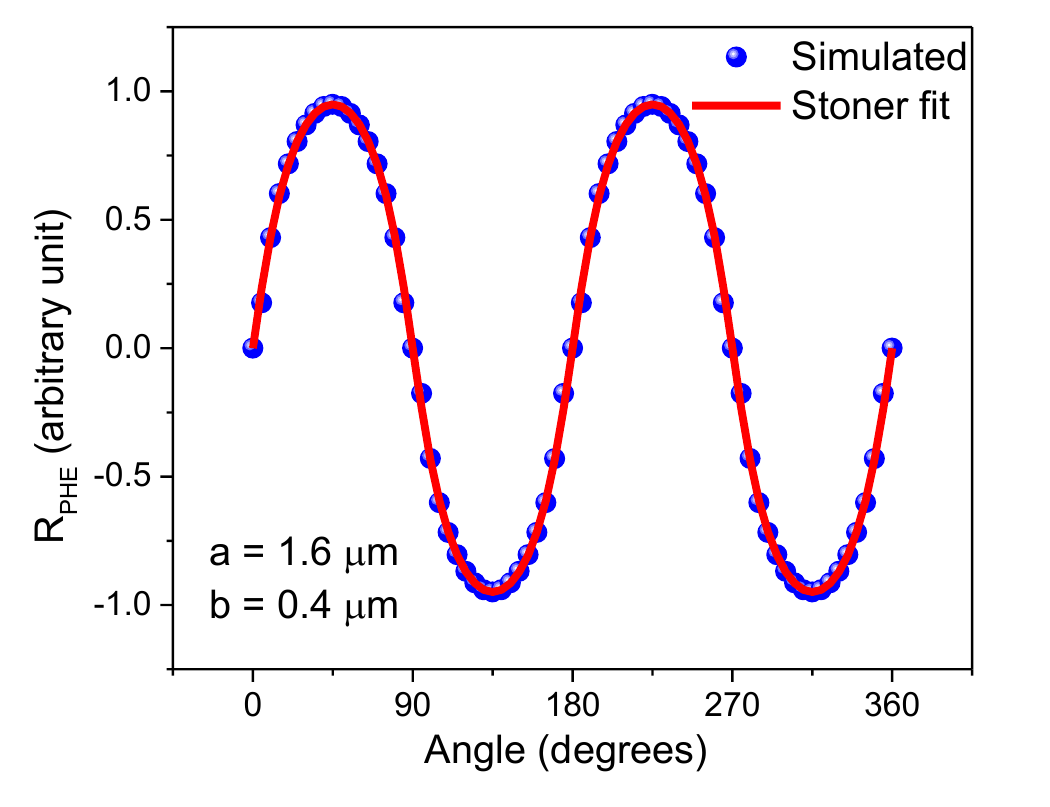}
    \caption{
    Representative angular-response fit used to extract the effective crossed-ellipse anisotropy field \(H_k\). The example corresponds to \(a=\SI{1.6}{\micro m}\) and \(b=\SI{0.4}{\micro m}\). The simulated field-on planar Hall response was calculated as \(R_{\mathrm{PHE}}=2\langle m_x\rangle\langle m_y\rangle\) and normalized for comparison. The fitted curve was obtained by minimizing the Stoner-type crossed-ellipse angular energy at each applied-field angle. The extracted \(H_k\) was then used as the effective ordinary anisotropy scale in the comparison \(H_s=H_0+C H_k\).
    }
    \label{fig:si_hk_fit_example}
\end{figure}

Figure~\ref{fig:si_hk_fit_example} shows a representative fit for the \(a=\SI{1.6}{\micro m}\), \(b=\SI{0.4}{\micro m}\) crossed ellipse. The fitted curve reproduces the angular dependence of the simulated field-on \(R_{\mathrm{PHE}}\), demonstrating that the simplified Stoner-type angular model captures the dominant ordinary anisotropy response used to extract \(H_k\).

The \(H_k\)-based comparison is applied only to the fixed-major-axis aspect-ratio series. In this case, changing \(a/b\) changes the ordinary in-plane anisotropy scale of the crossed ellipse, and the high-aspect-ratio OS branch follows an approximately linear relation of the form
\[
H_s = H_0 + C H_k .
\]
At lower aspect ratios, \(H_s\) deviates from this ordinary-anisotropy trend because the reversal becomes more nonuniform and the NS branch appears. Thus, the \(H_k\)-based comparison supports the interpretation that the low-aspect-ratio regime weakens the ordinary four-state confinement and enables a richer multistate remanent landscape.

This comparison is not intended to describe the fixed-\(a/b\) size-scaling series. In that case, the aspect ratio is held constant and the dominant change is the absolute magnetic length scale. Exchange, edge confinement, nonuniform reversal, and the increasing normalized thickness-to-width ratio \(c/b\) modify the switching pathway, so the size-scaling behavior is governed by finite-size micromagnetic effects rather than by the angular anisotropy scale alone.

\section{Aspect-ratio dependence without dead-layer correction}

\begin{figure}[H]
    \centering
    \includegraphics[width=0.5\textwidth, trim=0cm 9.2cm 0cm 0cm, clip]{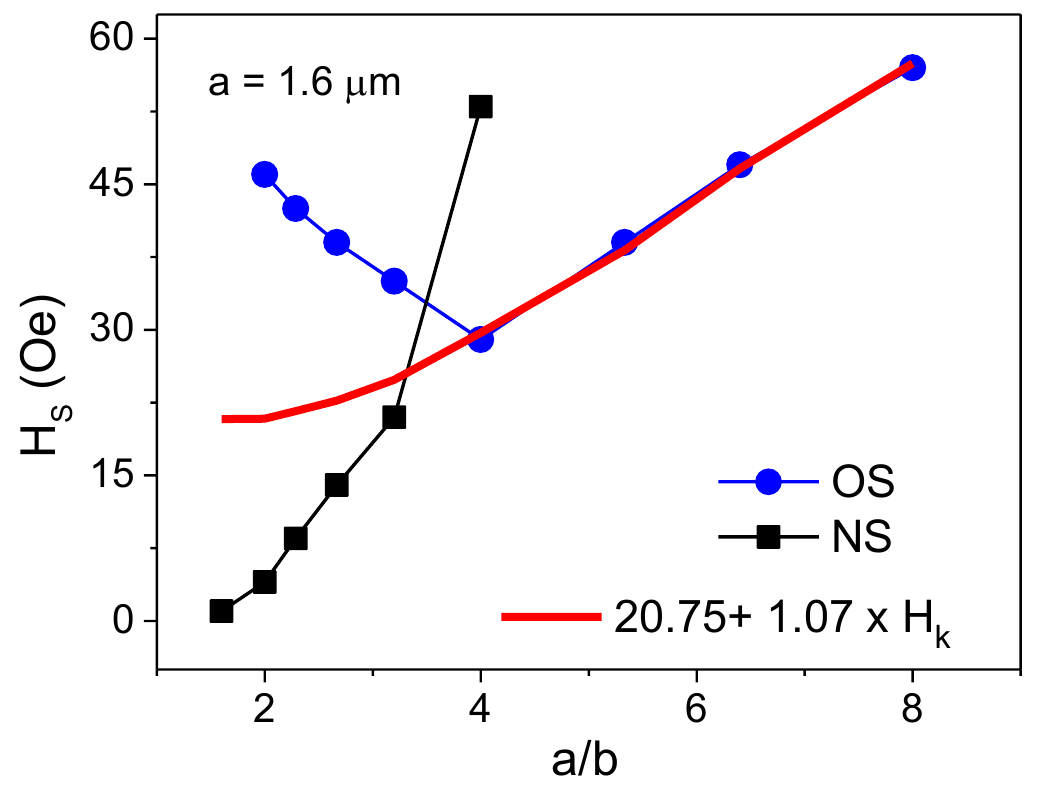}
    \caption{
    Aspect-ratio dependence of the switching field for the fixed-major-axis geometry \(a=\SI{1.6}{\micro m}\), calculated without including the magnetic dead-layer correction. This figure corresponds to the same geometry series as Fig.~3 in the main text, but here the full Py thickness \(t_{\mathrm{Py}}=\SI{2}{nm}\) was treated as magnetically active. The ordinary-state (OS) and newly stabilized-state (NS) branches remain present. The red curve shows the effective-anisotropy comparison \(H_s = \SI{20.75}{Oe} + 1.07 H_k\), where \(H_k\) was extracted from angular-response fitting using a Stoner-type crossed-ellipse model.
    }
    \label{fig:s2_aspect_ratio_no_dead_layer}
\end{figure}

Figure~\ref{fig:s2_aspect_ratio_no_dead_layer} shows the aspect-ratio-dependent switching field for the same fixed-major-axis geometry discussed in Fig.~3 of the main text, but calculated without including the magnetic dead-layer correction. In this supplementary case, the full Py thickness of \(t_{\mathrm{Py}}=\SI{2}{nm}\) was treated as magnetically active. Thus, the geometry series is directly comparable to the main-text aspect-ratio study, while isolating the effect of the effective magnetic thickness assumption.

The same qualitative behavior is preserved: the ordinary-state (OS) branch shows a nonmonotonic dependence on \(a/b\), with a low-field region at intermediate aspect ratio, while an additional newly stabilized-state (NS) branch appears at lower aspect ratios. The OS-branch trend was compared with an effective crossed-ellipse anisotropy field \(H_k\), extracted from angular-response fitting using a Stoner-type two-crossed-ellipse model. The red curve in Fig.~\ref{fig:s2_aspect_ratio_no_dead_layer} shows the corresponding linear comparison,
\[
H_s = \SI{20.75}{Oe} + 1.07 H_k .
\]
The no-dead-layer calculation therefore preserves the same branch structure and the same physical interpretation as the main-text calculation: the high-aspect-ratio OS branch is governed mainly by the ordinary crossed-ellipse anisotropy scale, while the low-aspect-ratio regime deviates as NS states become accessible.

\section{Aspect-ratio dependence for different major-axis sizes}

The main text focuses on aspect-ratio tuning for a fixed major axis \(a=\SI{1.6}{\micro m}\). To test whether the same qualitative behavior persists when the absolute lateral size is changed, additional aspect-ratio sweeps were performed for \(a=\SI{3.2}{\micro m}\) and \(a=\SI{0.8}{\micro m}\). These calculations show that the coexistence of OS and NS branches is a robust feature of the crossed-ellipse geometry, but that the switching-field scale, the location of the OS minimum, and the effective-anisotropy relation depend on the absolute major-axis size.

\subsection{Larger major axis, \texorpdfstring{\(a=\SI{3.2}{\micro m}\)}{a=3.2 µm}}

\begin{figure}[H]
    \centering
    \includegraphics[width=0.5\textwidth, trim=0cm 9.2cm 0cm 0cm, clip]{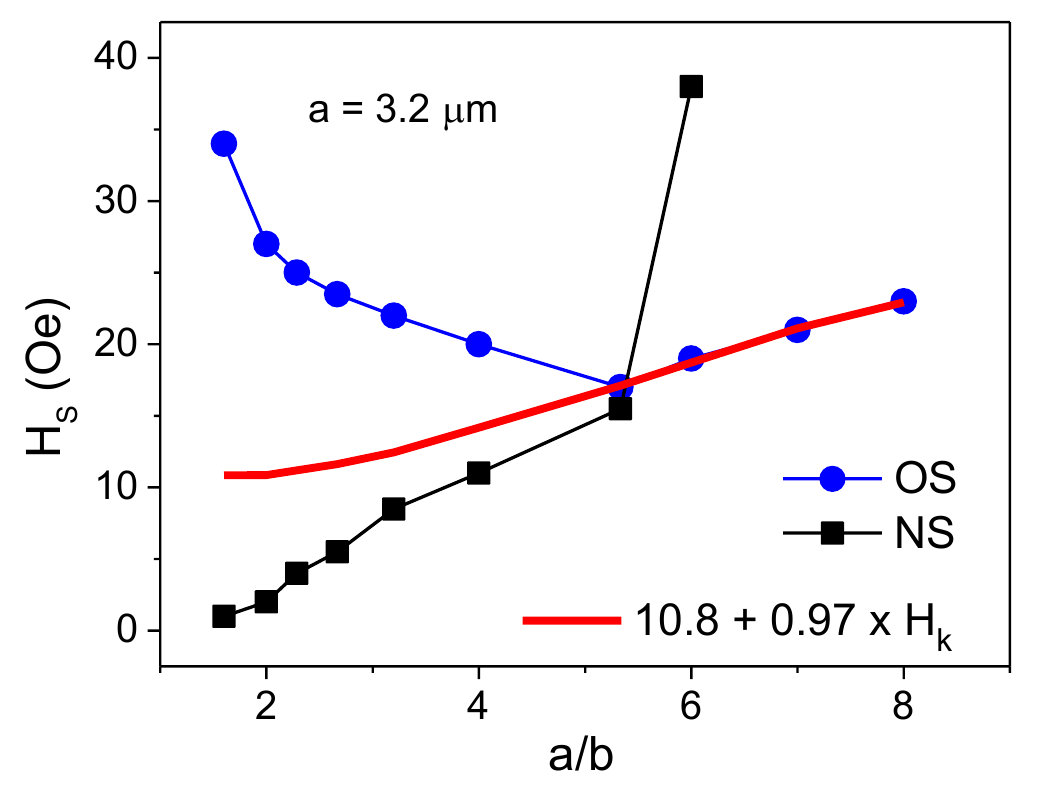}
    \caption{
    Aspect-ratio dependence of the switching field for a larger crossed-ellipse geometry with fixed major axis \(a=\SI{3.2}{\micro m}\). The ordinary-state (OS) branch shows a nonmonotonic dependence on \(a/b\), with a low-field region at intermediate aspect ratio. The newly stabilized-state (NS) branch appears at lower aspect ratios, showing that additional remanent-state branches also occur in the larger device. The red curve shows the effective-anisotropy comparison \(H_s=\SI{10.8}{Oe}+0.97H_k\), where \(H_k\) was extracted from angular-response fitting using a Stoner-type crossed-ellipse model.
    }
    \label{fig:s3_aspect_ratio_3p2um}
\end{figure}

Figure~\ref{fig:s3_aspect_ratio_3p2um} shows the aspect-ratio-dependent switching field for a larger crossed-ellipse geometry with fixed major axis \(a=\SI{3.2}{\micro m}\). The OS branch again shows a nonmonotonic dependence on \(a/b\), with a low-field region at intermediate aspect ratio. The NS branch also appears at lower aspect ratios, indicating that the emergence of additional remanent states is not limited to the smaller \(a=\SI{1.6}{\micro m}\) geometry. The OS branch was compared with the effective anisotropy field using
\[
H_s=\SI{10.8}{Oe}+0.97H_k .
\]
The reduced coefficient compared with the \(a=\SI{1.6}{\micro m}\) case indicates that the mapping between the effective angular anisotropy scale and the switching field depends on the absolute lateral size.

\subsection{Smaller major axis, \texorpdfstring{\(a=\SI{0.8}{\micro m}\)}{a=0.8 µm}}

\begin{figure}[H]
    \centering
    \includegraphics[width=0.5\textwidth, trim=0cm 9.2cm 0cm 0cm, clip]{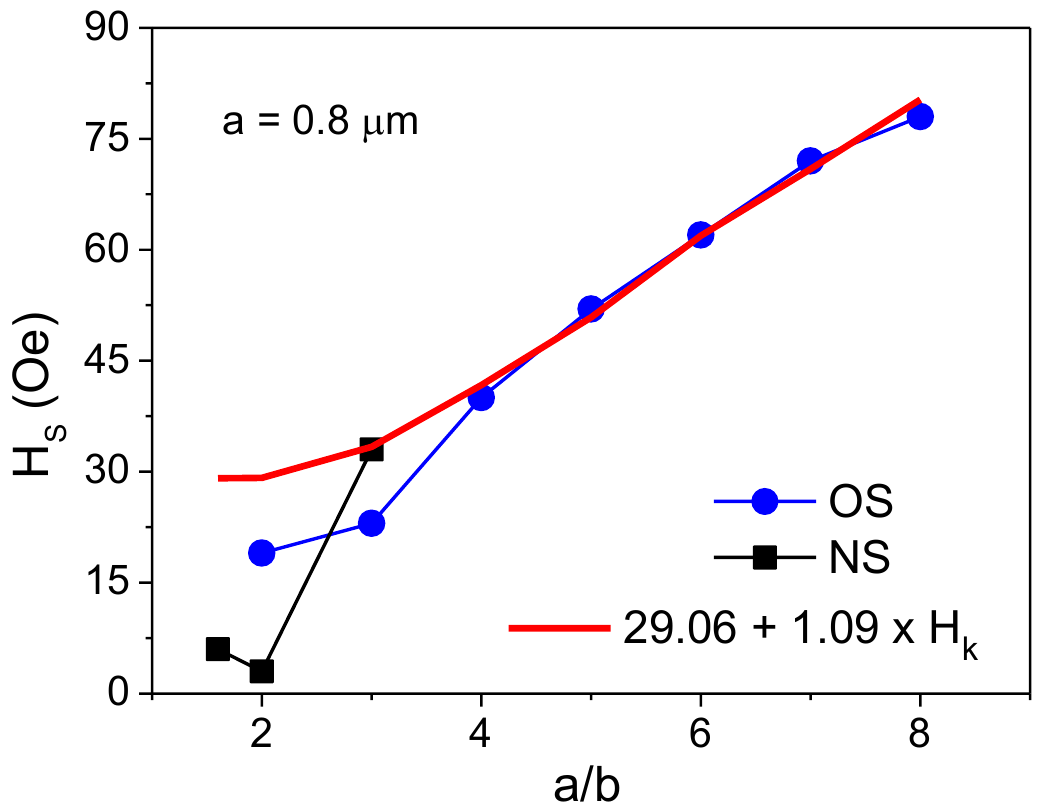}
    \caption{
    Aspect-ratio dependence of the switching field for a smaller crossed-ellipse geometry with fixed major axis \(a=\SI{0.8}{\micro m}\). The ordinary-state (OS) branch increases approximately monotonically with \(a/b\), unlike the \(a=\SI{1.6}{\micro m}\) case in the main text where a clearer intermediate-aspect-ratio minimum is observed. The newly stabilized-state (NS) branch appears at lower aspect ratios, indicating that additional remanent-state branches persist in the smaller geometry. The red curve shows the effective-anisotropy comparison \(H_s=\SI{29.06}{Oe}+1.09H_k\), where \(H_k\) was extracted from angular-response fitting using a Stoner-type crossed-ellipse model.
    }
    \label{fig:s4_aspect_ratio_0p8um}
\end{figure}

Figure~\ref{fig:s4_aspect_ratio_0p8um} shows the aspect-ratio-dependent switching field for a smaller crossed-ellipse geometry with fixed major axis \(a=\SI{0.8}{\micro m}\). Unlike the \(a=\SI{1.6}{\micro m}\) case in the main text, the OS branch for \(a=\SI{0.8}{\micro m}\) shows an approximately monotonic increase of \(H_s\) with increasing \(a/b\). This indicates that the optimum aspect-ratio behavior depends on the absolute lateral scale, not only on the aspect ratio itself. The NS branch is still observed at lower aspect ratios, showing that additional remanent-state branches persist even in the smaller geometry. The OS branch was compared with the effective anisotropy field using
\[
H_s=\SI{29.06}{Oe}+1.09H_k .
\]
The larger offset and coefficient compared with the larger devices are consistent with stronger finite-size confinement and a larger switching-field scale in the smaller geometry.

Taken together, the \(a=\SI{3.2}{\micro m}\), \(a=\SI{1.6}{\micro m}\), and \(a=\SI{0.8}{\micro m}\) aspect-ratio studies show that the OS/NS branch structure is a general feature of crossed ellipses. However, the exact aspect ratio that minimizes the OS switching field, the effective-anisotropy relation \(H_s=H_0+C H_k\), and the degree to which the NS branch is stabilized all depend on the absolute device size.

\section{Low-aspect-ratio state count and estimated MTJ resistance levels}

The main text shows that the \(1.6~\si{\micro m}\times0.8~\si{\micro m}\) device with \(a/b=2\) supports twelve field-off remanent states. To determine whether this low-aspect-ratio multistate response depends only on aspect ratio or also on absolute size, additional field-on/field-off angle sweeps were performed for two devices with the same \(a/b=2\), but different major axes: \(a=\SI{3.2}{\micro m}\), \(b=\SI{1.6}{\micro m}\), and \(a=\SI{0.8}{\micro m}\), \(b=\SI{0.4}{\micro m}\). In each case, the field-on configuration was obtained under an applied in-plane field of \(\SI{100}{Oe}\), and the corresponding field-off configuration was obtained after removing the field and relaxing the device to zero field. The simulated planar Hall response was calculated as \(R_{\mathrm{PHE}}=2\langle m_x\rangle\langle m_y\rangle\).

For comparison with the main-text MTJ estimate, each remanent state was also treated as a possible free-layer state of an in-plane-anisotropy MTJ. The projection onto an in-plane reference layer was calculated as \(q_i=\langle m_x\rangle_i\cos\phi_{\mathrm{ref}}+\langle m_y\rangle_i\sin\phi_{\mathrm{ref}}\). The corresponding resistance was estimated using \(R_i/R_P=(1+P)/(1+Pq_i)\), where \(P=\mathrm{TMR}/(\mathrm{TMR}+2)\). Resistance values were calculated for both \(\mathrm{TMR}=100\%\) and \(\mathrm{TMR}=200\%\). These values should be interpreted as ideal, reference-layer-dependent projection estimates rather than full device-level readout-margin calculations.

\subsection{Reference-layer-angle optimization for projected MTJ readout}

The projected MTJ resistance levels depend on the in-plane orientation of the reference layer. Therefore, the reference-layer angle \(\phi_{\mathrm{ref}}\) used in the resistance estimates was selected by scanning the minimum adjacent spacing between the calculated resistance levels. For each value of \(\phi_{\mathrm{ref}}\), the projected component \(q_i\) and corresponding normalized MTJ resistance were calculated for the selected field-off remanent states. The resistance levels were then sorted, and the smallest adjacent spacing was extracted.

\begin{figure}[H]
    \centering
    \includegraphics[width=0.5\textwidth, trim=0cm 9.2cm 0cm 0cm, clip]{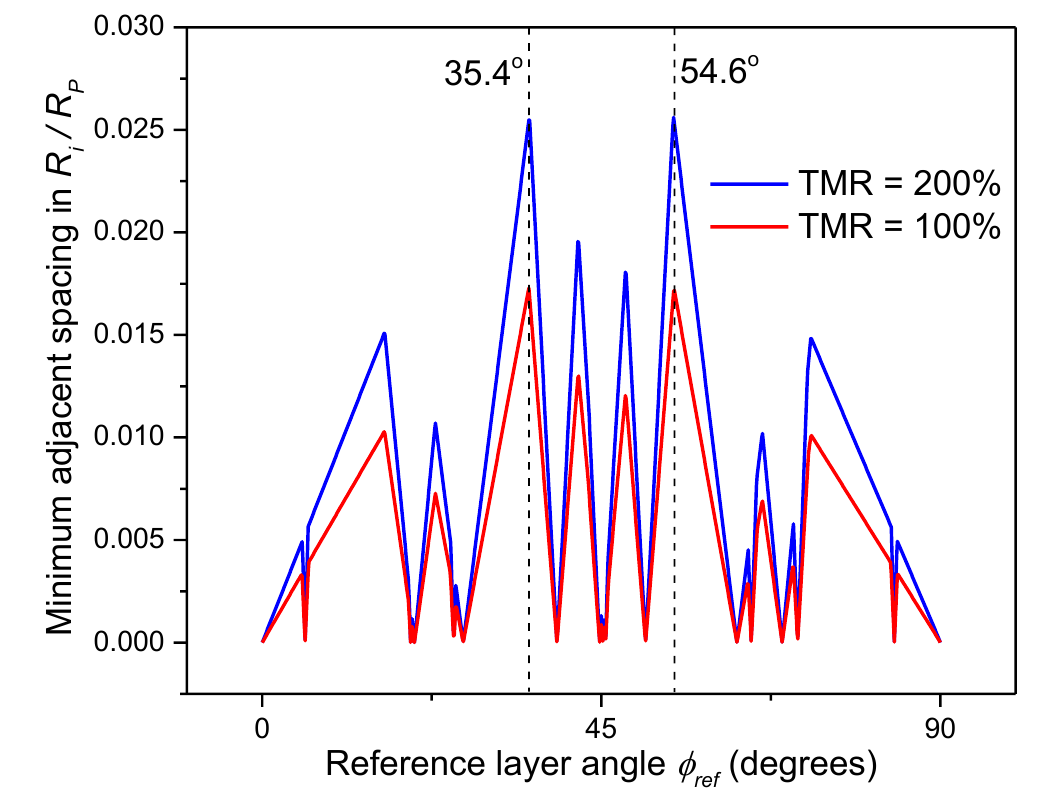}
    \caption{
    Reference-layer-angle dependence of the minimum adjacent projected MTJ resistance-level spacing for the twelve-state crossed-ellipse device. For each \(\phi_{\mathrm{ref}}\), the projected component \(q_i\) and resistance \(R_i/R_P\) were calculated for the selected field-off remanent states, the resistance levels were sorted, and the smallest adjacent spacing was extracted. The scan shows equivalent maxima near \(35.4^\circ\) and \(54.6^\circ\). The angle \(\phi_{\mathrm{ref}}=35.4^\circ\) was used in the main text and supplementary tables. The larger spacing obtained for \(\mathrm{TMR}=200\%\) shows that higher-TMR stacks can increase the projected resistance-level margin.
    }
    \label{fig:si_phi_ref_scan}
\end{figure}

Figure~\ref{fig:si_phi_ref_scan} shows the resulting minimum adjacent resistance-level spacing as a function of \(\phi_{\mathrm{ref}}\) for \(\mathrm{TMR}=100\%\) and \(\mathrm{TMR}=200\%\). The scan gives two equivalent maxima near \(35.4^\circ\) and \(54.6^\circ\), reflecting the approximate symmetry of the twelve-state remanent landscape. The value \(\phi_{\mathrm{ref}}=35.4^\circ\) was therefore used for the resistance estimates in the main text and in the supplementary tables. Increasing the assumed TMR from \(100\%\) to \(200\%\) increases the minimum adjacent projected level spacing, indicating that higher-TMR MTJ stacks can improve the readout margin for multistate operation.

\subsection{Twelve-state response for the larger \texorpdfstring{$3.2~\si{\micro m}\times1.6~\si{\micro m}$}{3.2 µm x 1.6 µm} device}

\begin{table}[H]
\centering
\small
\caption{Field-off remanent-state parameters and estimated MTJ resistance levels for the larger \(3.2~\si{\micro m}\times1.6~\si{\micro m}\) crossed-ellipse device at \(a/b=2\). The MTJ resistance estimate assumes an in-plane reference layer at \(\phi_{\mathrm{ref}}=35.4^\circ\).}
\label{tab:si_32um_mtj_states}
\begin{tabular}{cccccccc}
\toprule
State & Range & \(\phi_{\mathrm{OFF}}\) & \(R_{\mathrm{PHE}}\) & \(E_{\mathrm{tot}}\) & \(q_i\) & \multicolumn{2}{c}{\(R_i/R_P\)} \\
 & \((^\circ)\) & \((^\circ)\) &  & \((\mathrm{J\,m^{-3}})\) & & \(100\%\) TMR & \(200\%\) TMR \\
\midrule
S1  & \(5\text{--}25\)    & 11.9  &  0.317 & 141.30 &  0.8121 & 1.0493 & 1.0668 \\
S2  & \(30\text{--}60\)   & 45.0  &  0.782 & 166.72 &  0.8719 & 1.0331 & 1.0446 \\
S3  & \(65\text{--}85\)   & 78.1  &  0.317 & 141.30 &  0.6508 & 1.0957 & 1.1317 \\
S4  & \(95\text{--}115\)  & 101.9 & -0.317 & 141.30 &  0.3529 & 1.1930 & 1.2750 \\
S5  & \(120\text{--}150\) & 135.0 & -0.782 & 166.72 & -0.1474 & 1.4022 & 1.6193 \\
S6  & \(155\text{--}175\) & 168.1 & -0.317 & 141.30 & -0.6002 & 1.6668 & 2.1432 \\
S7  & \(185\text{--}205\) & 191.9 &  0.317 & 141.30 & -0.8121 & 1.8282 & 2.5255 \\
S8  & \(210\text{--}240\) & 225.0 &  0.782 & 166.72 & -0.8718 & 1.8796 & 2.6591 \\
S9  & \(245\text{--}265\) & 258.1 &  0.317 & 141.30 & -0.6510 & 1.7028 & 2.2239 \\
S10 & \(275\text{--}295\) & 281.9 & -0.317 & 141.30 & -0.3528 & 1.5110 & 1.8213 \\
S11 & \(300\text{--}330\) & 315.0 & -0.782 & 166.72 &  0.1474 & 1.2709 & 1.3970 \\
S12 & \(335\text{--}355\) & 348.1 & -0.317 & 141.30 &  0.6002 & 1.1110 & 1.1538 \\
\bottomrule
\end{tabular}
\end{table}

\begin{figure}[H]
    \centering
    \includegraphics[width=0.6\textwidth]{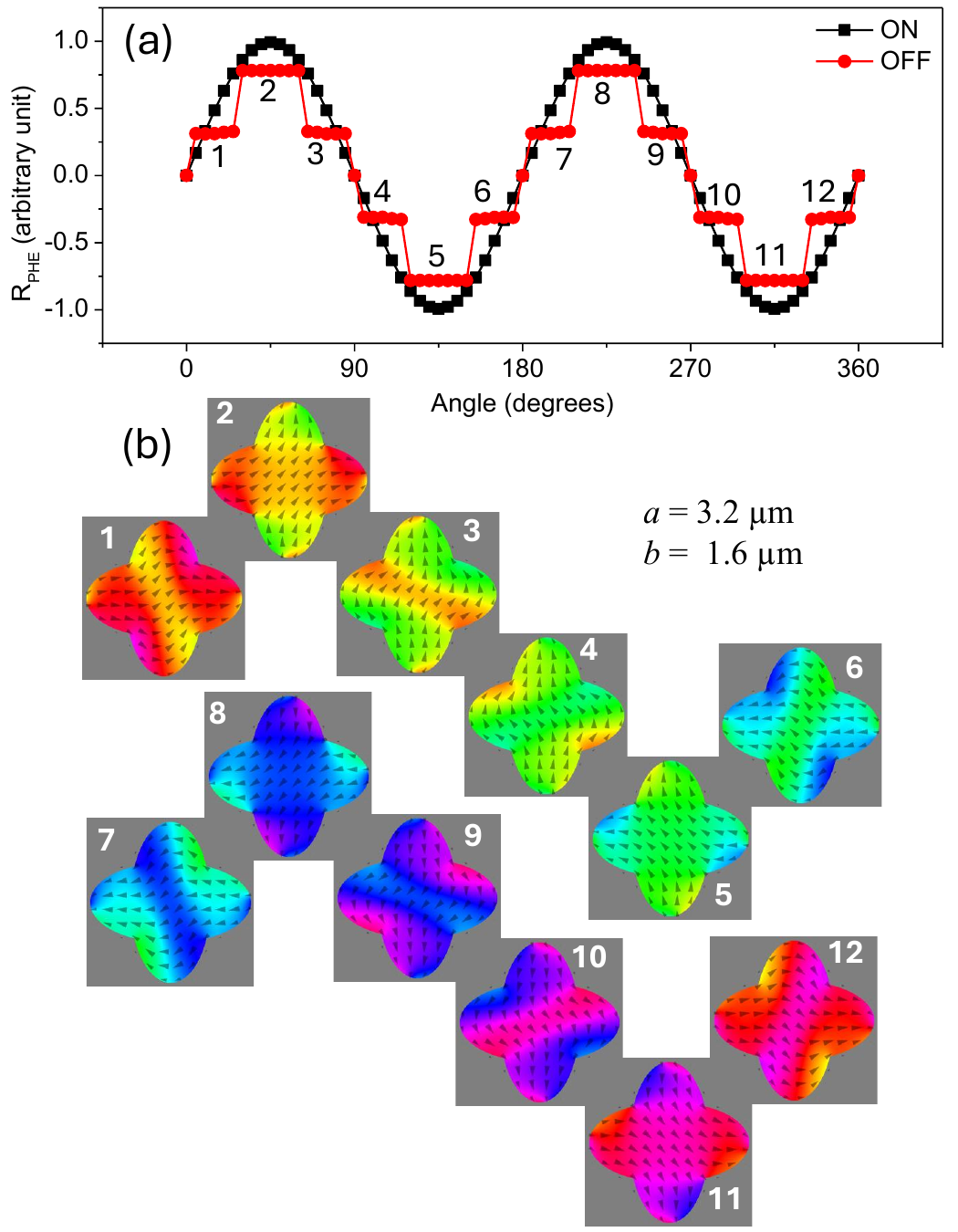}
    \caption{
    Angle-dependent field-on and field-off planar Hall response for the larger \(3.2~\si{\micro m}\times1.6~\si{\micro m}\) crossed-ellipse device with \(a/b=2\). Panel (a) shows the simulated \(R_{\mathrm{PHE}}\) response under field-on and field-off conditions at \(\SI{100}{Oe}\). Panel (b) shows the twelve field-off remanent configurations corresponding to the labeled plateaus in panel (a).
    }
    \label{fig:s5_32um_angle_sweep}
\end{figure}

For the larger \(3.2~\si{\micro m}\times1.6~\si{\micro m}\) device, the same \(a/b=2\) aspect ratio gives twelve well-resolved field-off remanent plateaus, as shown in Fig.~\ref{fig:s5_32um_angle_sweep} and summarized in Table~\ref{tab:si_32um_mtj_states}. The twelve states consist of eight canted states and four diagonal states. The exact principal-axis configurations at \(0^\circ\), \(90^\circ\), \(180^\circ\), and \(270^\circ\) occur only at the corresponding high-symmetry field angles and are therefore not counted as angular plateaus. Compared with the main-text \(1.6~\si{\micro m}\times0.8~\si{\micro m}\) device, the larger geometry preserves the twelve-state response and gives clearer separation of the planar Hall plateaus. This confirms that the twelve-state landscape is not unique to one device size.

\subsection{Eight-state response for the smaller \texorpdfstring{$0.8~\si{\micro m}\times0.4~\si{\micro m}$}{0.8 micron x 0.4 micron} device}

\begin{table}[H]
\centering
\small
\caption{Field-off remanent-state parameters and estimated MTJ resistance levels for the smaller \(0.8~\si{\micro m}\times0.4~\si{\micro m}\) crossed-ellipse device at \(a/b=2\). The MTJ resistance estimate assumes an in-plane reference layer at \(\phi_{\mathrm{ref}}=35.4^\circ\).}
\label{tab:si_08um_mtj_states}
\begin{tabular}{cccccccc}
\toprule
State & Range & \(\phi_{\mathrm{OFF}}\) & \(R_{\mathrm{PHE}}\) & \(E_{\mathrm{tot}}\) & \(q_i\) & \multicolumn{2}{c}{\(R_i/R_P\)} \\
 & \((^\circ)\) & \((^\circ)\) &  & \((\mathrm{J\,m^{-3}})\) & & \(100\%\) TMR & \(200\%\) TMR \\
\midrule
S1 & \(330\text{--}360,\,0\text{--}30\) & 0.0   &  0.000 & 632.86 &  0.7853 & 1.0567 & 1.0771 \\
S2 & \(35\text{--}55\)                  & 45.0  &  0.910 & 688.20 &  0.9407 & 1.0151 & 1.0202 \\
S3 & \(60\text{--}120\)                 & 90.0  &  0.000 & 632.84 &  0.5581 & 1.1242 & 1.1727 \\
S4 & \(125\text{--}145\)                & 135.0 & -0.910 & 688.20 & -0.1592 & 1.4080 & 1.6297 \\
S5 & \(150\text{--}210\)                & 180.0 &  0.000 & 632.86 & -0.7853 & 1.8061 & 2.4697 \\
S6 & \(215\text{--}235\)                & 225.0 &  0.910 & 688.20 & -0.9407 & 1.9424 & 2.8321 \\
S7 & \(240\text{--}300\)                & 270.0 &  0.000 & 632.84 & -0.5580 & 1.6380 & 2.0804 \\
S8 & \(305\text{--}325\)                & 315.0 & -0.910 & 688.20 &  0.1592 & 1.2662 & 1.3894 \\
\bottomrule
\end{tabular}
\end{table}

\begin{figure}[H]
    \centering
    \includegraphics[width=0.6\textwidth, trim=0cm 8.75cm 0cm 0cm, clip]{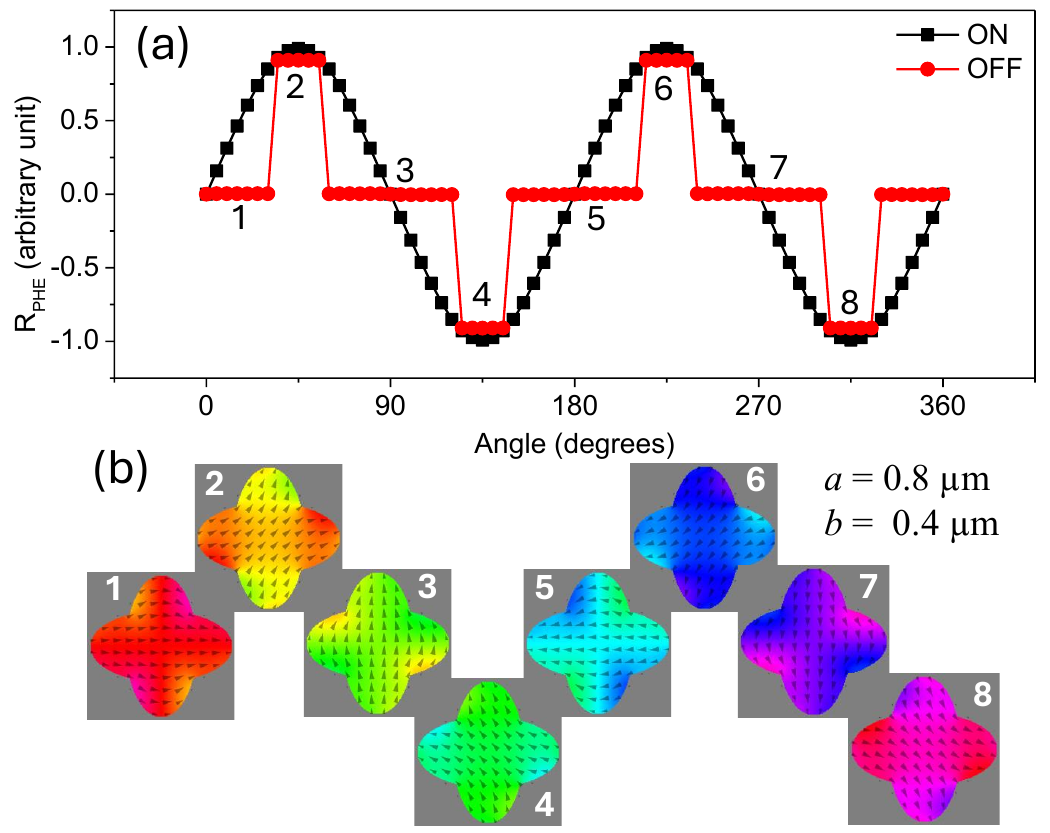}
    \caption{
    Angle-dependent field-on and field-off planar Hall response for the smaller \(0.8~\si{\micro m}\times0.4~\si{\micro m}\) crossed-ellipse device with \(a/b=2\). Panel (a) shows the simulated \(R_{\mathrm{PHE}}\) response under field-on and field-off conditions at \(\SI{100}{Oe}\). Panel (b) shows the eight field-off remanent configurations corresponding to the labeled plateaus in panel (a).
    }
    \label{fig:s6_08um_angle_sweep}
\end{figure}

For the smaller \(0.8~\si{\micro m}\times0.4~\si{\micro m}\) device, the same \(a/b=2\) aspect ratio does not preserve the twelve-state splitting. Instead, the field-off response gives eight well-resolved remanent plateaus, as shown in Fig.~\ref{fig:s6_08um_angle_sweep} and summarized in Table~\ref{tab:si_08um_mtj_states}. These states are located near the principal axes and diagonal directions. The principal-axis states have \(R_{\mathrm{PHE}}\approx0\) and lower energy density, \(E_{\mathrm{tot}}\approx\SI{632.85}{J/m^3}\), whereas the diagonal states have larger \(|R_{\mathrm{PHE}}|\) and higher energy density, \(E_{\mathrm{tot}}\approx\SI{688.20}{J/m^3}\). Thus, at this smaller lateral size, the near-axis states do not split into two separate canted states around each axis.

The comparison between Tables~\ref{tab:si_32um_mtj_states} and~\ref{tab:si_08um_mtj_states} shows that the low-aspect-ratio multistate behavior is not determined by \(a/b\) alone. At \(a/b=2\), the larger \(3.2~\si{\micro m}\) device supports twelve well-resolved field-off remanent states, whereas the smaller \(0.8~\si{\micro m}\) device supports eight. Together with the \(1.6~\si{\micro m}\times0.8~\si{\micro m}\) result in the main text, this indicates that the twelve-state landscape exists over an intermediate-to-large lateral-size range, while the smallest device studied here collapses the canted near-axis pairs into single axis-like states. Therefore, a full geometry classification must consider both aspect ratio and absolute size.

\subsection{State-count phase map}

\begin{figure}[H]
    \centering
    \includegraphics[width=0.7\textwidth]{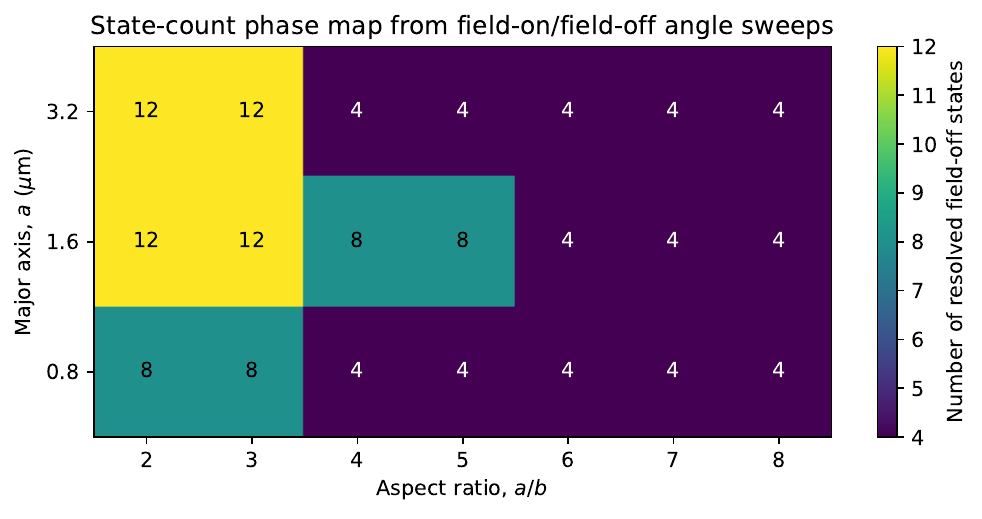}
    \caption{
    State-count phase map obtained from angle-dependent field-on/field-off sweeps. The color scale gives the number of resolved field-off remanent states as a function of major axis \(a\) and aspect ratio \(a/b\). High-aspect-ratio geometries primarily support four ordinary remanent states, whereas lower-aspect-ratio geometries stabilize additional states.
    }
    \label{fig:si_phase_space_map}
\end{figure}

Figure~\ref{fig:si_phase_space_map} summarizes the number of resolved field-off remanent states obtained from angle-dependent field-on/field-off sweeps for \(a=\SI{0.8}{\micro m}\), \(1.6~\si{\micro m}\), and \(3.2~\si{\micro m}\), with aspect ratios \(a/b=2\)--8. The state count was extracted from the discrete field-off plateaus in the planar Hall response and the corresponding remanent configurations.

A remanent state was counted as a robust resolved state only when the same field-off configuration appeared at two or more consecutive field-angle points in the \(\SI{5}{^\circ}\)-step angle sweep. Single-angle configurations that disappeared after a \(\SI{5}{^\circ}\) change in field direction were not counted as robust plateaus, even if their magnetization texture resembled one of the twelve-state configurations. This criterion avoids classifying isolated switching-angle events as stable angular plateaus.

The map shows that high-aspect-ratio devices mainly support the four ordinary remanent states, while lower-aspect-ratio devices support additional robust plateaus. Some intermediate-aspect-ratio sweeps produced isolated twelve-state-like configurations at single field angles, but these were not counted as robust states because they switched to an ordinary state after a \(\SI{5}{^\circ}\) change in field direction.

\subsection{Estimated MTJ resistance-level spacing}

\begin{figure}[H]
    \centering
    \includegraphics[width=0.7\textwidth, trim=0cm 8cm 0cm 0cm, clip]{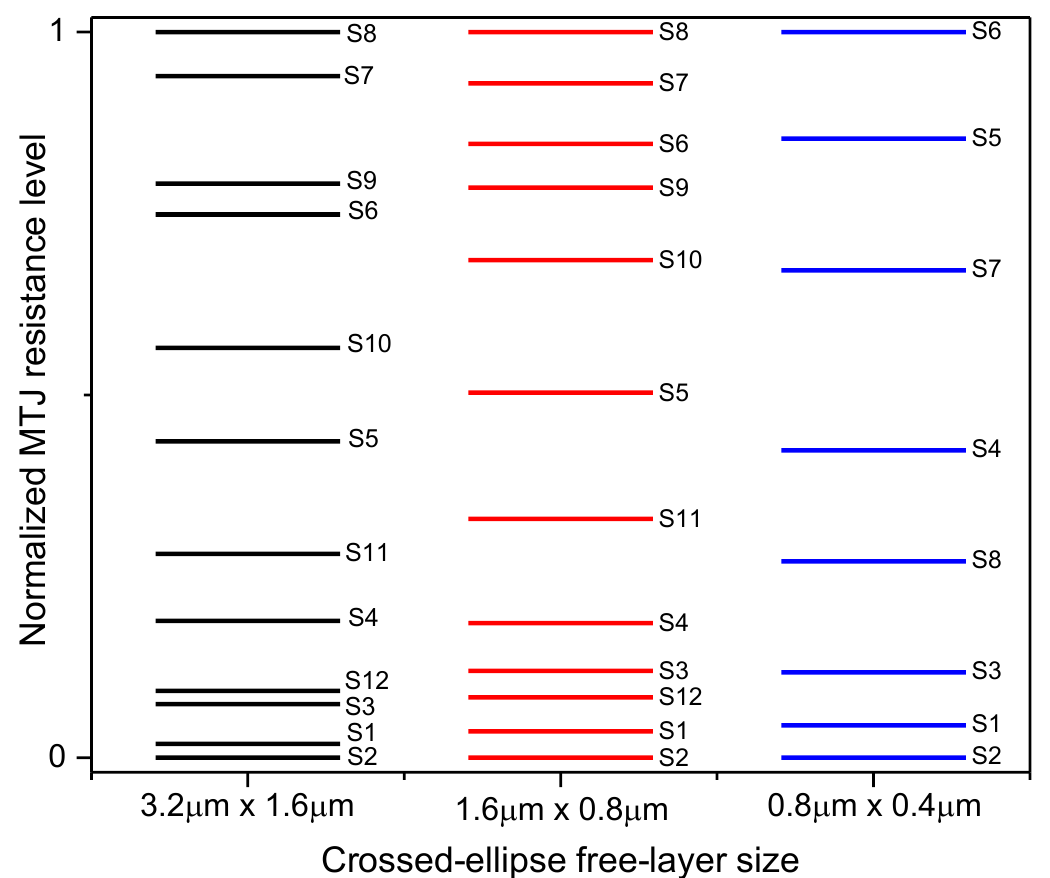}
    \caption{
    Estimated normalized MTJ resistance levels for low-aspect-ratio crossed-ellipse free layers with \(a/b=2\). The levels were calculated from the projection of each field-off remanent state onto an in-plane reference layer using the angular tunneling model described in the main text, and were normalized separately for each device size to compare level spacing. The \(3.2~\si{\micro m}\times1.6~\si{\micro m}\) and \(1.6~\si{\micro m}\times0.8~\si{\micro m}\) devices show twelve calculated resistance levels, whereas the \(0.8~\si{\micro m}\times0.4~\si{\micro m}\) device shows eight levels. Closely spaced levels indicate that practical multilevel readout will require sufficient TMR ratio, low resistance noise, and optimized reference-layer orientation.
    }
    \label{fig:si_mtj_resistance_levels}
\end{figure}

Figure~\ref{fig:si_mtj_resistance_levels} compares the estimated MTJ resistance levels for the three low-aspect-ratio crossed-ellipse devices with \(a/b=2\). The resistance values were calculated from the projection of each field-off remanent state onto an in-plane reference layer using the same angular tunneling model described above. The plotted values are normalized separately for each device size to compare relative level spacing. The corresponding numerical resistance estimates for both \(\mathrm{TMR}=100\%\) and \(\mathrm{TMR}=200\%\) are listed in the tables above.

The \(3.2~\si{\micro m}\times1.6~\si{\micro m}\) and \(1.6~\si{\micro m}\times0.8~\si{\micro m}\) devices show twelve calculated resistance levels, consistent with their twelve field-off remanent states. In contrast, the smaller \(0.8~\si{\micro m}\times0.4~\si{\micro m}\) device shows eight calculated resistance levels, consistent with the reduced number of remanent states in this geometry. Several levels are closely spaced, especially in the twelve-state devices. Therefore, although the magnetic states can be mapped into multiple projected MTJ resistance levels, practical multilevel readout will require sufficient tunneling magnetoresistance, low resistance noise, and an optimized reference-layer orientation.

\section{String-method estimates of inter-state energy barriers}

To estimate representative inter-state barriers, zero-field string-method calculations were performed for selected transitions between remanent states of the \(1.6~\si{\micro m}\times0.8~\si{\micro m}\) crossed-ellipse device. The approach follows the general string-method idea of relaxing a discretized path between two metastable states while reparameterizing the path coordinate, and is closely related to micromagnetic minimum-energy-path and NEB approaches used for magnetic energy-barrier calculations \cite{eRenVandenEijndenString,eRenVandenEijndenSimplifiedString,dittrichPathMethod,naganumaStringMTJ}.

\subsection{String-method procedure}

The relaxed field-off magnetization configurations from the twelve-state angular sweep were used as the end points of each path. The calculations were performed at zero applied field and zero current. Each path was initialized by normalized cell-by-cell interpolation of the full micromagnetic configuration between the two end states. The use of the full magnetization texture is important because the remanent states are nonuniform and cannot be represented reliably by the spatially averaged magnetization direction alone.

For each transition, a discrete string consisting of 24 images was used, including the two fixed end states, following the standard chain-of-images representation used in string and NEB-type minimum-energy-path calculations \cite{eRenVandenEijndenString,dittrichPathMethod}. The intermediate images were relaxed iteratively using MuMax3. During each string iteration, the end states were kept fixed, while the intermediate images were evolved for a short high-damping zero-field relaxation step. The resulting images were then reparameterized to maintain approximately equal spacing along the path coordinate. This relax--reparameterize cycle was repeated until the maximum energy along the path changed slowly or until the path clearly relaxed to a barrierless connection.

After each iteration, the energy of every image along the reparameterized path was evaluated at zero field. The forward and reverse barriers were calculated as
\[
E_b^{i\rightarrow j}=\max(E_{\mathrm{path}})-E_i
\]
and
\[
E_b^{j\rightarrow i}=\max(E_{\mathrm{path}})-E_j ,
\]
where \(E_i\) and \(E_j\) are the endpoint energies. The corresponding thermal stability factor was estimated as
\[
\Delta=\frac{E_b}{k_{\mathrm{B}}T},
\]
using \(T=\SI{300}{K}\). These calculations should be interpreted as representative minimum-energy-path estimates for the relevant symmetry classes of the twelve-state landscape, rather than as a full device-level retention analysis including disorder, temperature-dependent dynamics, and stack-specific material variations.

\subsection{Barrier hierarchy among selected remanent states}

Three representative transitions were examined. The S3--S4 transition connects two adjacent canted states near the \(+y\) axis. These states have nearly equal relaxed energies and the smallest angular separation among the twelve plateaus, making this pair a natural candidate for the lowest barrier. The S2--S3 transition connects a diagonal state to a neighboring canted state and therefore probes the barrier between the two energy-density families. The S2--S5 transition connects two diagonal states across the upper branch of the angular landscape and probes a more strongly separated transition within the diagonal-state family.

\begin{table}[H]
\centering
\small
\caption{
Representative zero-field string-method barrier estimates for selected
transitions between field-off remanent states of the
\(1.6~\si{\micro m}\times0.8~\si{\micro m}\) crossed-ellipse device.
The thermal stability factor was calculated as
\(\Delta=E_b/k_{\mathrm{B}}T\) at \(T=\SI{300}{K}\).
The S3--S4 and S2--S3 calculations connect neighbouring remanent
configurations and therefore provide direct information on local escape
barriers. The S2--S5 calculation connects more widely separated endpoints
and should instead be interpreted as a representative direct-path barrier;
the physically relevant transition may proceed through intermediate minima
with smaller individual barriers. Consequently, the values in this table
establish the hierarchy of selected inter-state barriers but are not by
themselves sufficient to determine the total number of thermally independent
memory states.}

\label{tab:si_string_barriers}
\begin{tabular}{lcccc}
\toprule
Transition & Type & \(E_b^{\mathrm{forward}}\) (J) & \(\Delta_{\mathrm{forward}}\) & Interpretation \\
\midrule
S3\(\rightarrow\)S4 & canted--canted & \(\approx 0\) & \(\approx 0\) & barrierless/merged basin \\
S2\(\rightarrow\)S3 & diagonal--canted & \(1.58\times10^{-19}\) & 38.2 & finite barrier \\
S3\(\rightarrow\)S2 & canted--diagonal & \(3.64\times10^{-19}\) & 87.8 & finite barrier \\
S2\(\rightarrow\)S5 & diagonal--diagonal & \(5.40\times10^{-19}\) & 130.3 & large symmetric barrier \\
S5\(\rightarrow\)S2 & diagonal--diagonal & \(5.40\times10^{-19}\) & 130.3 & large symmetric barrier \\
\bottomrule
\end{tabular}
\end{table}

\begin{figure}[H]
    \centering
    \includegraphics[width=0.8\textwidth, trim=0cm 12.5cm 0cm 0.5cm, clip]{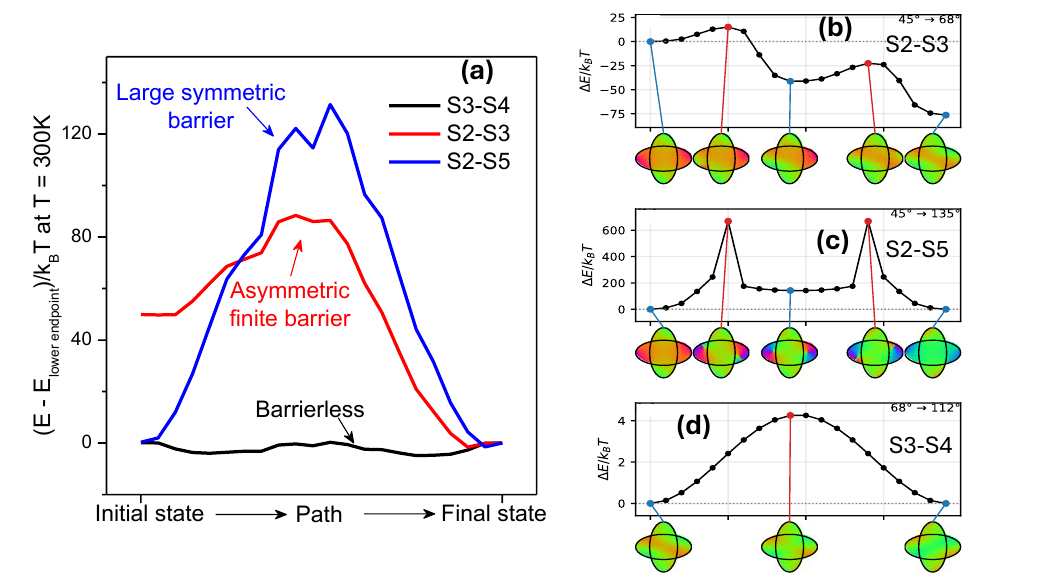}
    \caption{Representative zero-field minimum-energy-path calculations for selected
    transitions between remanent states of the
    \(1.6~\si{\micro m}\times0.8~\si{\micro m}\) crossed-ellipse device.
    Panel (a) shows the MuMax3 string-method energy profiles, with the energy
    referenced to the lower-energy endpoint and normalized by
    \(k_{\mathrm{B}}T\) at \(T=\SI{300}{K}\), i.e.
    \((E-E_{\mathrm{lower~endpoint}})/k_{\mathrm{B}}T\).
    The S3--S4 canted-pair path is essentially barrierless, the S2--S3
    diagonal-to-canted transition shows an asymmetric finite barrier, and the
    direct S2--S5 diagonal-to-diagonal path shows a substantially larger
    nearly symmetric barrier.
    Panels (b)--(d) show independent MuMax+ minimum-energy-path calculations
    for the corresponding S2--S3, S2--S5, and S3--S4 transitions,
    respectively. The independent calculations reproduce the very small
    barrier between the closest canted states and further show that the
    physically relevant transition between more widely separated
    configurations can involve intermediate minima rather than a single
    direct endpoint-to-endpoint barrier. The combined MuMax3 and MuMax+
    calculations therefore establish a strongly nonuniform inter-state
    barrier landscape and show that thermal retention must be evaluated from
    the lowest neighbouring-state escape pathways.}
    \label{fig:si_string_barriers}
\end{figure}

Representative relaxed string energy profiles are shown in Fig.~\ref{fig:si_string_barriers}, and the corresponding barrier values are summarized in Table~\ref{tab:si_string_barriers}. The S3--S4 calculation shows that the initially interpolated path contains a small apparent energy maximum, but repeated string relaxation lowers the maximum until it reaches the endpoint energy within numerical resolution. Thus, the S3--S4 pair is effectively barrierless in the relaxed minimum-energy-path calculation. This indicates that the closest canted states are not thermally independent nonvolatile memory states.

In contrast, the S2--S3 path retains a substantial finite barrier after string relaxation. After 100 string iterations, the forward S2\(\rightarrow\)S3 barrier is \(E_b\approx1.58\times10^{-19}~\mathrm{J}\), corresponding to \(\Delta\approx38\), while the reverse S3\(\rightarrow\)S2 barrier is \(E_b\approx3.64\times10^{-19}~\mathrm{J}\), corresponding to \(\Delta\approx88\). The asymmetry arises because S2 is a higher-energy diagonal-family state, whereas S3 is a lower-energy canted-family state.

The S2--S5 transition gives a still larger barrier. Because S2 and S5 are symmetry-related diagonal states with nearly equal endpoint energies, the forward and reverse barriers are nearly identical. After 100 string iterations, the barrier remains \(E_b\approx5.4\times10^{-19}~\mathrm{J}\), corresponding to \(\Delta\approx130\). Therefore, diagonal-state transitions are strongly separated compared with the closest canted-pair transition.

An important limitation is that a string calculation between two
non-neighbouring remanent configurations does not necessarily determine the
lowest physical escape barrier between those states. If one or more
metastable minima lie between the selected endpoints, the minimum-energy
transition route can consist of a sequence of neighbouring-state transitions
rather than a single direct path. Independent MuMax+ calculations explicitly
show this behavior for the present twelve-state landscape: a direct path
between more widely separated configurations can yield a large apparent
barrier, while a cascade through intermediate canted states contains
substantially smaller individual barriers. Consequently, direct
endpoint-to-endpoint barriers are interpreted here as representative
energy-landscape comparisons unless the selected states are confirmed to be
neighbouring minima connected through a single saddle point.

The barrier hierarchy therefore provides an important interpretation
of the twelve-state angular response. The twelve field-off plateaus are
reproducibly accessible under the field-on/field-off protocol, but they do not
all correspond to independent thermally stable memory levels. The closest
canted states, such as S3 and S4, are separated by a very small barrier and
therefore cannot be regarded as distinct nonvolatile states at room
temperature. Larger barriers obtained for selected diagonal-to-canted and
diagonal-to-diagonal endpoint pairs demonstrate that the energy landscape
contains more strongly separated configurations, but these direct barriers do
not by themselves establish the retention of the corresponding states. The
relevant retention barrier for each remanent configuration is instead set by
the lowest available neighbouring-state escape pathway.

On this basis, the present calculations establish twelve accessible
remanent configurations together with a strongly nonuniform hierarchy of
thermal barriers, but they do not uniquely establish eight nonvolatile
basins. Determining the complete thermally independent state count would
require a connected minimum-energy-path network among neighbouring minima,
or complementary finite-temperature dynamical simulations. The present
\(1.6~\si{\micro m}\times0.8~\si{\micro m}\) geometry should therefore be
viewed as a demonstration of a geometry-generated multistate energy landscape
rather than as an already optimized nonvolatile multilevel memory cell.
Its thermal robustness can be improved through further optimization of aspect
ratio, magnetic thickness, anisotropy, material parameters, and stack design.

\section{CoFeB parameter check for transferability of the geometry trends}

\begin{table}[H]
\centering
\small
\caption{
MuMax3 parameters used for the CoFeB transferability checks. The crossed-ellipse geometry, thickness, cell size, field protocol, and analysis procedure were kept the same as in the Py simulations, while the saturation magnetization and exchange stiffness were changed to representative CoFeB-like values.
}
\label{tab:si_cofeb_mumax_parameters}
\begin{tabular}{lll}
\toprule
Parameter & Symbol / MuMax3 quantity & Value used \\
\midrule
Material & -- & CoFeB-like ferromagnet \\
Saturation magnetization & \(M_s\) / \texttt{Msat} & \(1.3\times10^{6}~\mathrm{A\,m^{-1}}\) \\
Exchange stiffness & \(A_{\mathrm{ex}}\) / \texttt{Aex} & \(1.5\times10^{-11}~\mathrm{J\,m^{-1}}\) \\
Gilbert damping & \(\alpha\) / \texttt{alpha} & 0.01 \\
Magnetic thickness & \(t_{\mathrm{FM}}\) & \(\SI{1.7}{nm}\) \\
Cell size & \(\Delta x,\Delta y,\Delta z\) & \(\SI{5}{nm},\SI{5}{nm},\SI{2}{nm}\) \\
Edge smoothing & \texttt{EdgeSmooth} & 8 \\
Fixed major axis for aspect-ratio scan & \(a\) & \(\SI{1.6}{\micro m}\) \\
Minor axis range & \(b\) & varied to set \(a/b\) \\
Low-aspect-ratio multistate geometry & \(a\times b\) & \(\SI{1.6}{\micro m}\times\SI{0.8}{\micro m}\) \\
Applied field for angular sweep & \(H\) & \(\SI{100}{Oe}\) \\
Field protocol & -- & field-on relaxation followed by zero-field relaxation \\
Planar Hall proxy & \(R_{\mathrm{PHE}}\) & \(2\langle m_x\rangle\langle m_y\rangle\) \\
\bottomrule
\end{tabular}
\end{table}

The main simulations in this work use Py crossed-ellipse free layers as a well-characterized micromagnetic model system. Practical high-TMR MTJs, however, are more commonly based on CoFeB/MgO/CoFeB stacks. To test whether the geometry-dependent trends are specific to Py or persist for a more MTJ-relevant ferromagnet, additional MuMax3 simulations were performed using representative CoFeB parameters.

For these calculations, the saturation magnetization and exchange stiffness were changed to CoFeB-like values, while the crossed-ellipse geometry, thickness, field protocol, and analysis procedure were kept the same as in the Py simulations. The purpose of this comparison is not to provide a complete CoFeB device optimization, but to test whether the ordinary-state (OS) and newly stabilized-state (NS) branch structure, as well as the low-aspect-ratio twelve-state response, remain present when the magnetic material parameters are changed.

\subsection{Aspect-ratio dependence of CoFeB crossed ellipses}

\begin{figure}[H]
    \centering
    \includegraphics[width=0.58\textwidth, trim=0cm 9.2cm 0cm 0cm, clip]{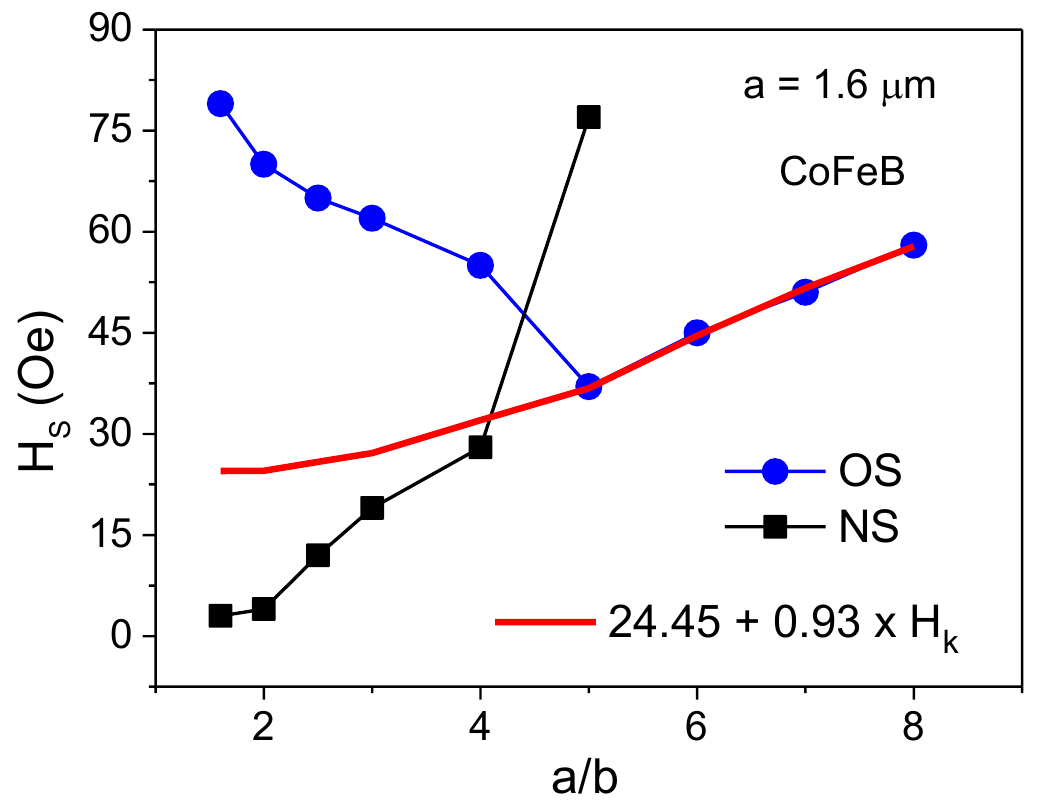}
    \caption{
    Aspect-ratio dependence of the switching field for CoFeB crossed ellipses with fixed major axis \(a=\SI{1.6}{\micro m}\). The ordinary-state (OS) and newly stabilized-state (NS) branches remain present, showing that the branch structure is mainly geometry-controlled. The red curve shows the effective-anisotropy comparison \(H_s=\SI{24.45}{Oe}+0.93H_k\), where \(H_k\) was extracted from angular-response fitting using a Stoner-type crossed-ellipse model.
    }
    \label{fig:si_cofeb_aspect_ratio}
\end{figure}

Figure~\ref{fig:si_cofeb_aspect_ratio} shows the aspect-ratio dependence of the switching field for CoFeB crossed ellipses with fixed major axis \(a=\SI{1.6}{\micro m}\). As in the Py case, the aspect ratio was varied by changing the minor axis \(b\), while the same field-relaxation protocol was used to identify the ordinary-state (OS) and newly stabilized-state (NS) branches.

The CoFeB calculation preserves the qualitative branch structure observed for Py. The OS branch appears at higher aspect ratios and shows an increase in switching field as \(a/b\) increases, while the NS branch remains accessible at lower aspect ratios. Following the updated analysis used for the Py aspect-ratio figures, the OS-branch switching field was compared with the effective crossed-ellipse anisotropy field \(H_k\), extracted from angular-response fitting using a Stoner-type crossed-ellipse model. For CoFeB, the OS branch follows the approximate relation
\[
H_s=\SI{24.45}{Oe}+0.93H_k .
\]
The larger offset compared with the Py \(a=\SI{1.6}{\micro m}\) case reflects the higher switching-field scale of the CoFeB-like material parameters, while the persistence of the \(H_k\)-based trend indicates that the high-aspect-ratio OS branch remains governed by the ordinary crossed-ellipse anisotropy scale.

This comparison indicates that the OS/NS branch structure and the effective-anisotropy trend are primarily controlled by the crossed-ellipse geometry, whereas the absolute switching-field scale is material-dependent. Therefore, the Py simulations in the main text can be interpreted as a geometry-design map, while CoFeB/MgO provides a natural material platform for future high-TMR implementation.

\subsection{Angle-dependent twelve-state response for CoFeB at \texorpdfstring{$a/b=2$}{a/b=2}}

\begin{figure}[H]
    \centering
    \includegraphics[width=0.7\textwidth]{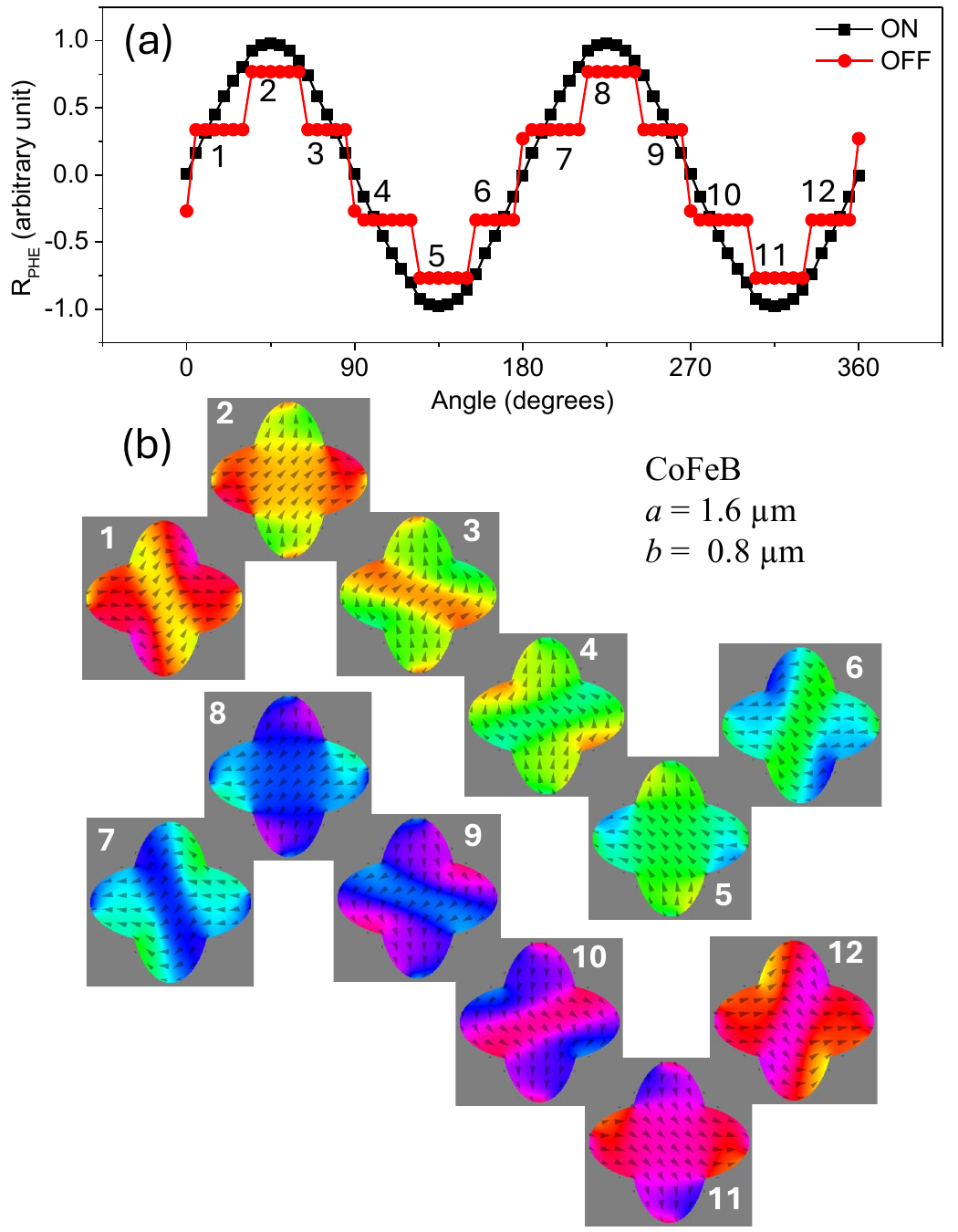}
    \caption{
    Angle-dependent field-on and field-off response for the CoFeB crossed-ellipse device with \(a=\SI{1.6}{\micro m}\), \(b=\SI{0.8}{\micro m}\), and \(a/b=2\). Panel (a) shows the simulated planar Hall response, calculated as \(R_{\mathrm{PHE}}=2\langle m_x\rangle\langle m_y\rangle\), under field-on and field-off conditions. The field-off response separates into twelve discrete remanent plateaus. Panel (b) shows the corresponding relaxed field-off remanent configurations. The persistence of twelve remanent states indicates that the low-aspect-ratio multistate landscape is retained for CoFeB-like material parameters.
    }
    \label{fig:si_cofeb_angle_sweep}
\end{figure}

The low-aspect-ratio CoFeB device with \(a=\SI{1.6}{\micro m}\) and \(b=\SI{0.8}{\micro m}\) was tested using the same field-on/field-off angle-sweep protocol used for the Py twelve-state analysis. The field-on configuration was obtained under an applied in-plane field of \(\SI{100}{Oe}\), and the field-off configuration was obtained after removing the field and relaxing the magnetization to zero field. The planar Hall response was calculated as \(R_{\mathrm{PHE}}=2\langle m_x\rangle\langle m_y\rangle\).

Figure~\ref{fig:si_cofeb_angle_sweep} shows that the CoFeB crossed ellipse also supports twelve discrete field-off remanent plateaus at \(a/b=2\). The field-on response follows the applied field direction, whereas the field-off response separates into discrete remanent states after relaxation. The corresponding remanent-state maps confirm that these states are nonuniform configurations stabilized by the crossed-ellipse geometry. Thus, the twelve-state low-aspect-ratio landscape is retained when the material parameters are changed from Py-like to CoFeB-like values.

These CoFeB checks support the interpretation that the state-count enhancement is a geometry-driven effect. The larger magnetic moment of CoFeB modifies the absolute switching-field and energy scales, but the OS/NS branch structure and the low-aspect-ratio twelve-state response persist. This is important for MTJ applications because CoFeB/MgO stacks can provide substantially larger tunneling magnetoresistance than Py-based junctions, improving the projected electrical readout margin for multistate operation.

\section{Preliminary array-coupling check}

As a first estimate of inter-cell magnetostatic coupling in an array environment, a 3\(\times\)3 crossed-ellipse array was simulated and the center cell was analyzed. The surrounding cells were included to provide a nearest-neighbor magnetostatic environment, and the center-cell switching field was compared with the corresponding isolated-cell value.

\begin{figure}[H]
    \centering
    \includegraphics[width=0.9\textwidth, trim=0cm 13cm 0cm 0cm, clip]{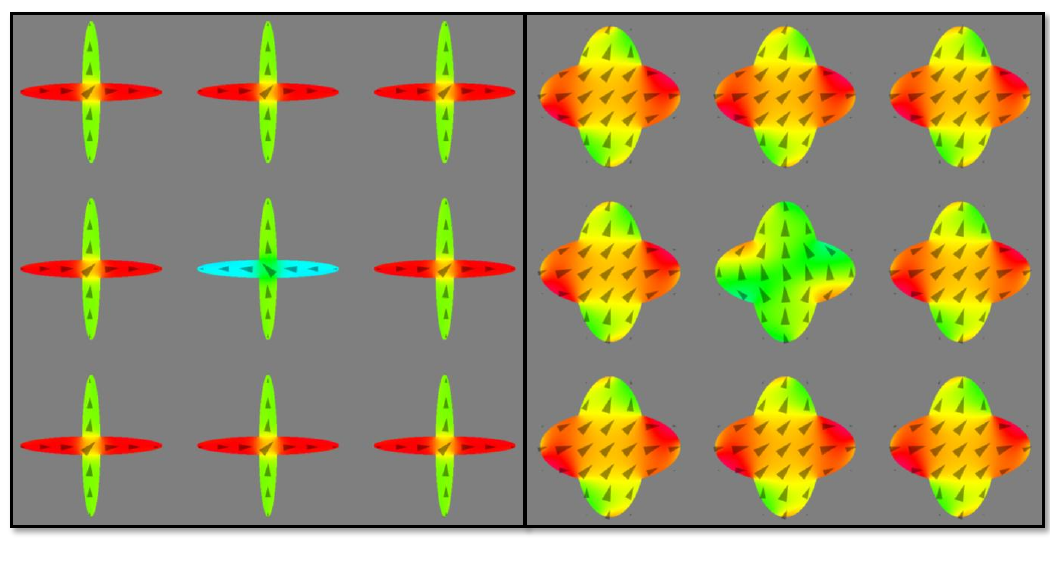}
    \caption{
    Representative 3\(\times\)3 crossed-ellipse arrays used for the preliminary magnetostatic pitch-coupling check. The center cell, shown in a different color, is the measured cell, while the surrounding cells provide the nearest-neighbor magnetostatic environment. The high-aspect-ratio \(a/b=8\) ordinary-state geometry shows only a small switching-field change, from \(\SI{42.2}{Oe}\) in the isolated-cell simulation to \(\SI{42.6}{Oe}\) in the array at \(p/a=1.25\). The low-aspect-ratio newly stabilized-state branch is more sensitive, changing from \(\SI{2.9}{Oe}\) to \(\SI{4.5}{Oe}\) at \(p/a=1.25\), and to \(\SI{3.6}{Oe}\) at \(p/a=2.5\).
    }
    \label{fig:si_array_coupling}
\end{figure}

Figure~\ref{fig:si_array_coupling} shows representative 3\(\times\)3 array configurations, and Table~\ref{tab:si_array_coupling} summarizes the corresponding switching-field changes for the measured center cell. For the high-aspect-ratio \(a=\SI{1.6}{\micro m}\), \(b=\SI{0.2}{\micro m}\), \(a/b=8\) ordinary-state branch at \(p/a=1.25\), the switching field changes only from \(\SI{42.2}{Oe}\) for the isolated cell to \(\SI{42.6}{Oe}\) in the array, corresponding to a relative change of approximately \(0.95\%\). This indicates that nearest-neighbor magnetostatic coupling is weak for the strongly shape-confined four-state geometry at the tested pitch.

The lower-aspect-ratio newly stabilized-state branch is more sensitive in relative terms. For this branch, the switching field changes from \(\SI{2.9}{Oe}\) in the isolated-cell calculation to \(\SI{4.5}{Oe}\) at \(p/a=1.25\). Increasing the pitch to \(p/a=2.5\) reduces the array switching field to \(\SI{3.6}{Oe}\), indicating that the magnetostatic perturbation weakens with increasing pitch. Although the relative change remains larger than for the high-aspect-ratio ordinary-state branch, the absolute switching field remains much smaller than that of the high-aspect-ratio branch. This behavior is consistent with the weaker ordinary anisotropy confinement of the low-aspect-ratio multistate geometry.

\begin{table}[H]
\centering
\small
\caption{
Preliminary 3\(\times\)3 array-coupling check. The switching field of the measured center cell is compared with the corresponding isolated-cell value. The relative change is calculated as \((H_s^{\mathrm{array}}-H_s^{\mathrm{isolated}})/H_s^{\mathrm{isolated}}\times100\%\).
}
\label{tab:si_array_coupling}
\begin{tabular}{lccccc}
\toprule
Geometry / branch & \(p/a\) & \(H_s^{\mathrm{isolated}}\) & \(H_s^{\mathrm{array}}\) & \(\Delta H_s\) & Relative change \\
 &  & (Oe) & (Oe) & (Oe) & (\%) \\
\midrule
\(a/b=8\), OS  & 1.25 & 42.2 & 42.6 & 0.4 & 0.95 \\
Low-\(a/b\), NS & 1.25 & 2.9  & 4.5  & 1.6 & 55.2 \\
Low-\(a/b\), NS & 2.50 & 2.9  & 3.6  & 0.7 & 24.1 \\
\bottomrule
\end{tabular}
\end{table}

These preliminary calculations indicate that the high-aspect-ratio ordinary-state branch is only weakly altered by nearest-neighbor magnetostatic coupling at the tested pitch. The low-aspect-ratio multistate branch is more sensitive, but the perturbation decreases when the pitch is increased from \(p/a=1.25\) to \(p/a=2.5\), and the absolute switching field remains low. This check does not replace a full crossbar analysis, since write-line selectivity, Oersted-field cross-talk, current distribution, Joule heating, and readout-margin variations require separate treatment.

\section{Bias-field check for reference-layer-induced asymmetry}

\begin{figure}[H]
    \centering
    \includegraphics[width=0.7\textwidth]{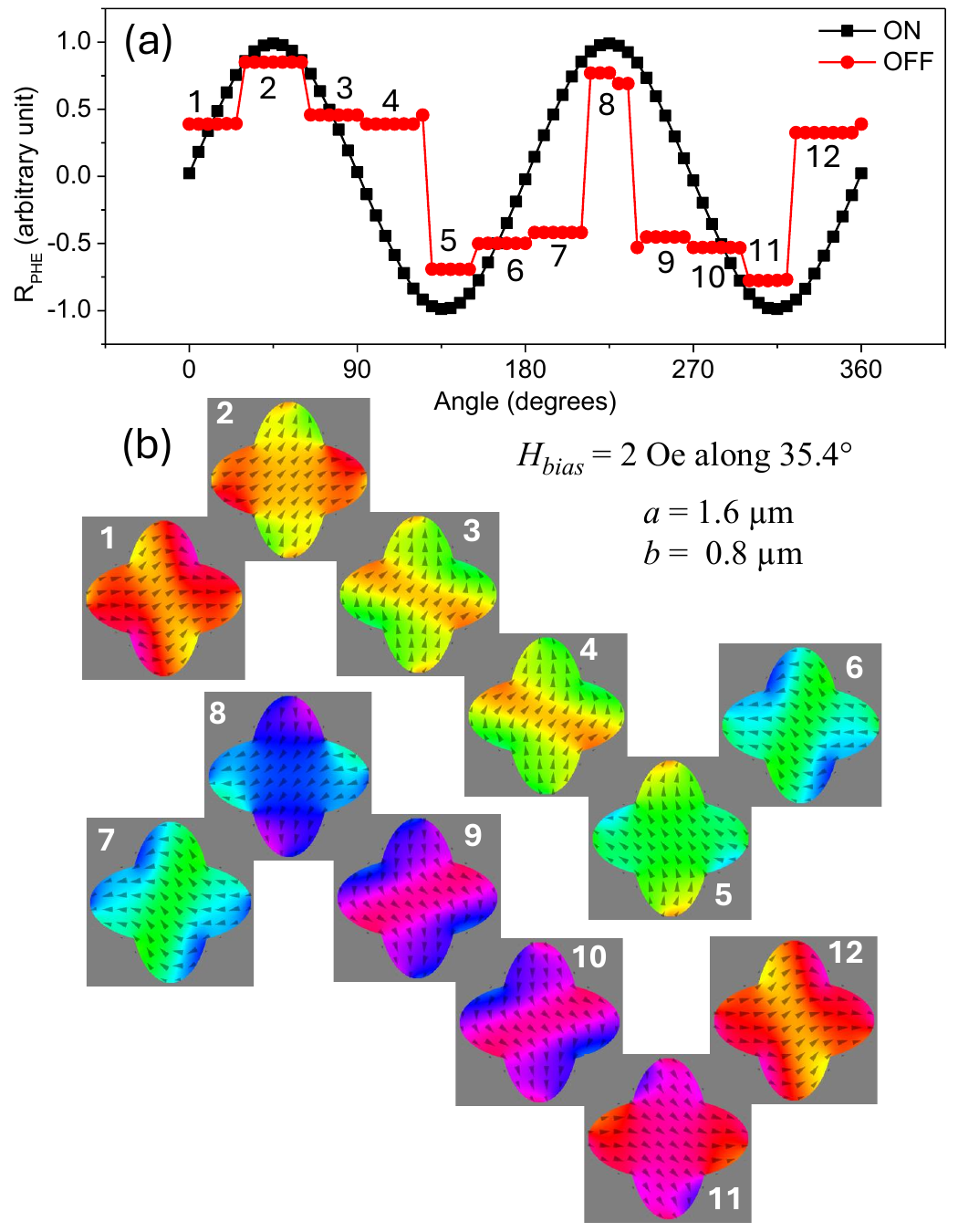}
    \caption{
    Bias-field check for the \(a=\SI{1.6}{\micro m}\), \(b=\SI{0.8}{\micro m}\), \(a/b=2\) crossed-ellipse device with \(H_{\mathrm{bias}}=\SI{2}{Oe}\) applied along \(\phi_{\mathrm{ref}}=35.4^\circ\). Panel (a) shows the simulated planar Hall response, and panel (b) shows the corresponding field-off remanent configurations. The ordinary diagonal states remain distinguishable, while the closest newly stabilized canted states begin to merge, especially the S3--S4 pair.
    }
    \label{fig:si_bias_field_2oe}
\end{figure}

In the idealized crossed-ellipse simulations, the free layer is treated as an isolated magnetic element. In a full MTJ stack, however, the reference layer can produce a stray field or weak coupling field at the free layer. Such a field can break the approximate angular symmetry of the crossed-ellipse energy landscape and modify the number of accessible remanent states. As a first proxy for this effect, additional field-on/field-off angular sweeps were performed for the \(a=\SI{1.6}{\micro m}\), \(b=\SI{0.8}{\micro m}\), \(a/b=2\) Py crossed ellipse with static in-plane bias fields applied along the optimized reference-layer direction, \(\phi_{\mathrm{ref}}=35.4^\circ\). In practical MTJ stacks, this bias can be reduced by using a small centered reference ellipse and further minimized by using a synthetic-antiferromagnetic reference stack, which compensates the net dipolar field at the free layer. The uniform-bias simulations below should therefore be viewed as conservative sensitivity tests for uncompensated reference-layer stray field.

\begin{figure}[H]
    \centering
    \includegraphics[width=0.7\textwidth]{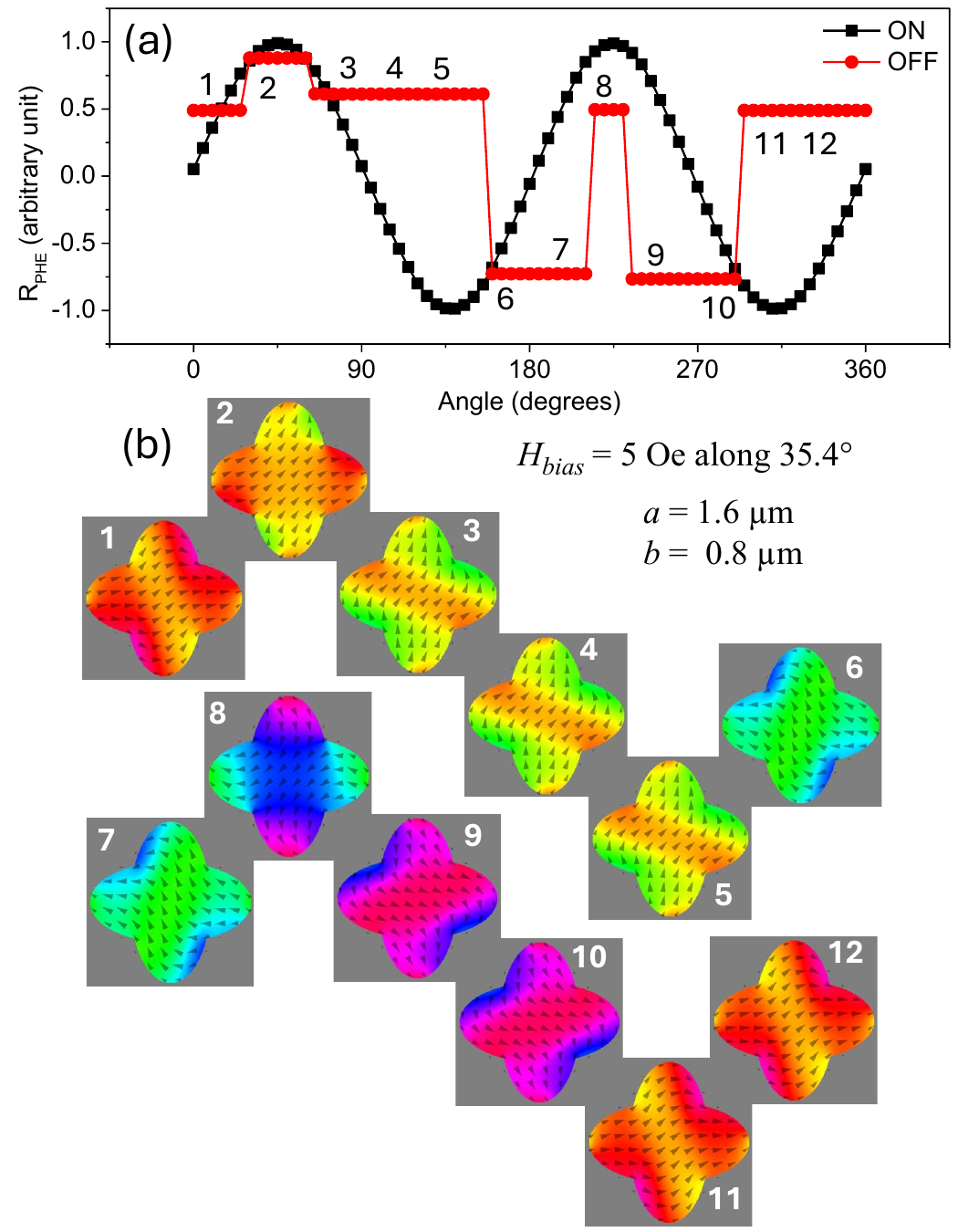}
    \caption{
    Bias-field check for the \(a=\SI{1.6}{\micro m}\), \(b=\SI{0.8}{\micro m}\), \(a/b=2\) crossed-ellipse device with \(H_{\mathrm{bias}}=\SI{5}{Oe}\) applied along \(\phi_{\mathrm{ref}}=35.4^\circ\). Compared with the unbiased twelve-configuration response, the bias field breaks the approximate angular symmetry and merges several neighboring configurations. States 3--4--5, 6--7, 9--10, and 11--12--1 form common field-off plateaus, while states 2 and 8 remain distinct, giving six accessible field-off states under this biased condition.
    }
    \label{fig:si_bias_field_5oe}
\end{figure}

The bias field was included during both the field-on and field-off relaxation steps. During the field-on step, the rotating applied field was superposed with the static bias field. During the field-off step, the rotating field was removed, while the static bias field was retained. This procedure tests whether the low-aspect-ratio multistate response survives a reference-layer-like symmetry-breaking perturbation.

\begin{figure}[H]
    \centering
    \includegraphics[width=0.7\textwidth]{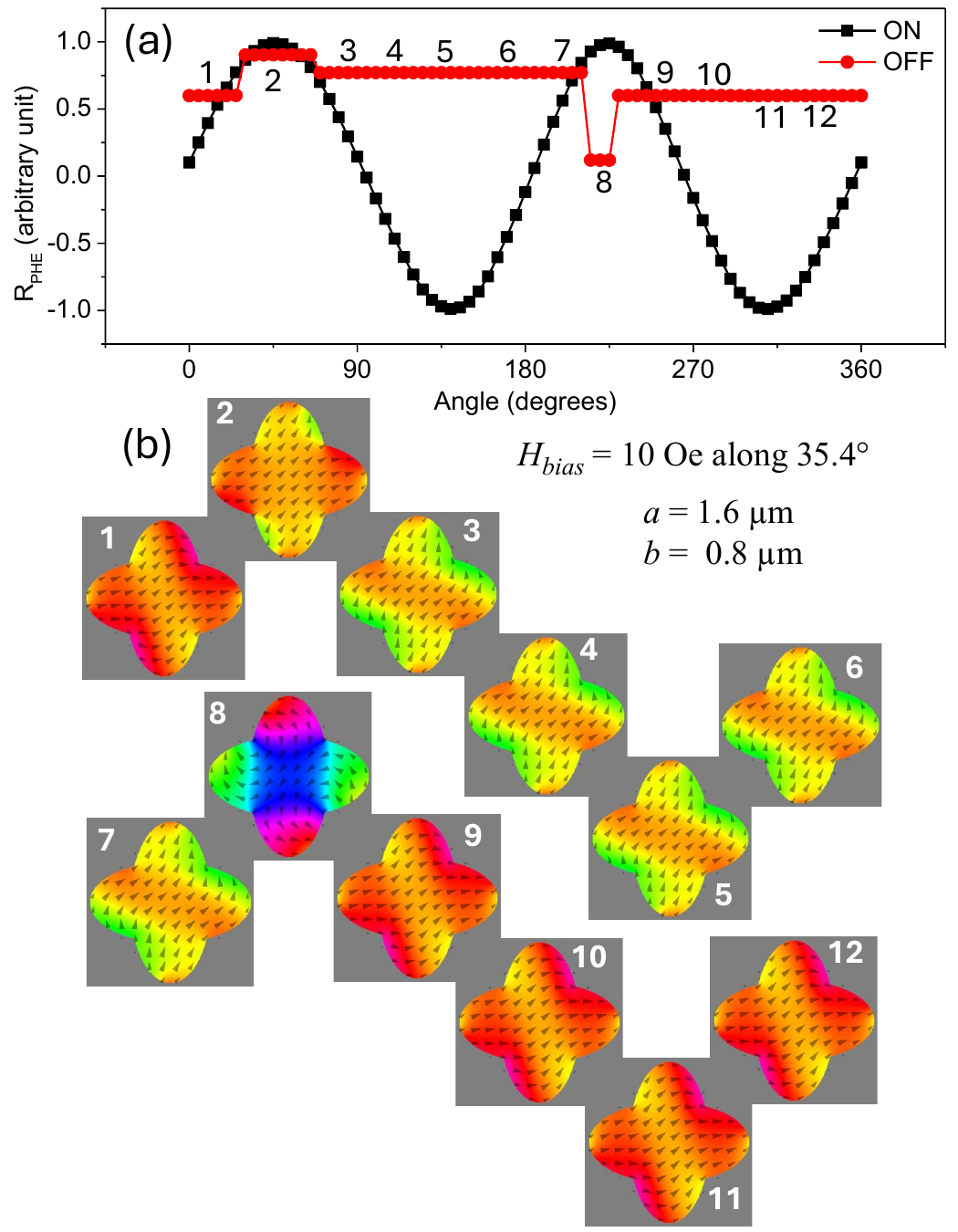}
    \caption{
    Bias-field check for the \(a=\SI{1.6}{\micro m}\), \(b=\SI{0.8}{\micro m}\), \(a/b=2\) crossed-ellipse device with \(H_{\mathrm{bias}}=\SI{10}{Oe}\) applied along \(\phi_{\mathrm{ref}}=35.4^\circ\). The stronger bias field substantially collapses the angular state landscape: states 3--4--5--6--7 merge into one broad field-off plateau, and states 9--10--11--12--1 merge into another. This indicates that stronger reference-layer-like bias fields can strongly reduce the effective accessible state count.
    }
    \label{fig:si_bias_field_10oe}
\end{figure}

Figures~\ref{fig:si_bias_field_2oe}--\ref{fig:si_bias_field_10oe} show the field-on/field-off response for \(H_{\mathrm{bias}}=\SI{2}{Oe}\), \(\SI{5}{Oe}\), and \(\SI{10}{Oe}\), respectively. The results are summarized in Table~\ref{tab:si_bias_field_check}. For the weak-bias case, \(H_{\mathrm{bias}}=\SI{2}{Oe}\), the ordinary diagonal states remain distinct, while the closest newly stabilized canted pairs begin to merge, most notably the S3--S4 pair. Thus, a weak reference-layer-like bias perturbs the fine canted-state splitting but does not destroy the multistate landscape.

For \(H_{\mathrm{bias}}=\SI{5}{Oe}\), the symmetry breaking becomes stronger and several neighboring configurations merge. In particular, states 3--4--5, 6--7, 9--10, and 11--12--1 form common field-off plateaus. Together with the distinct states 2 and 8, this gives six accessible field-off states under the biased condition. Therefore, the low-aspect-ratio crossed ellipse remains multistate, but the exact number of accessible states is reduced relative to the unbiased twelve-configuration response.

\begin{table}[H]
\centering
\small
\caption{
Bias-field check for the \(a=\SI{1.6}{\micro m}\), \(b=\SI{0.8}{\micro m}\), \(a/b=2\) crossed-ellipse device. A static in-plane bias field was applied along the optimized reference-layer direction, \(\phi_{\mathrm{ref}}=35.4^\circ\), during both field-on and field-off relaxation. The table summarizes the main merging behavior observed in the field-off response.
}
\label{tab:si_bias_field_check}
\begin{tabular}{cll}
\toprule
\(H_{\mathrm{bias}}\) & Main observed behavior & Effective accessible response \\
(Oe) &  &  \\
\midrule
0  & Twelve accessible field-off configurations & 12 configurations \\
2  & OS states remain distinct; closest NS pairs begin to merge, especially S3--S4 & Multistate response largely retained \\
5  & S3--S4--S5, S6--S7, S9--S10, and S11--S12--S1 merge & 6 accessible states \\
10 & S3--S4--S5--S6--S7 and S9--S10--S11--S12--S1 merge; OS states are strongly affected & Strongly collapsed multistate response \\
\bottomrule
\end{tabular}
\end{table}

For the stronger-bias case, \(H_{\mathrm{bias}}=\SI{10}{Oe}\), the ordinary-state plateaus are also strongly affected. States 3--4--5--6--7 merge into one broad field-off plateau, and states 9--10--11--12--1 merge into another broad plateau, while states 2 and 8 remain distinguishable. This shows that sufficiently strong reference-layer-like bias can substantially collapse the angular state landscape.

These bias-field checks show that the additional low-aspect-ratio states are not solely an artifact of exact fourfold symmetry, because multistate behavior survives under weak and moderate bias fields. However, the practical state count is sensitive to the magnitude and direction of the bias field. Therefore, the final state count in a full MTJ stack will depend not only on the crossed-ellipse free-layer geometry, but also on reference-layer stray field, interlayer coupling, and stack design.

\section{\space Sensitivity to lithographic edge roughness}

To test whether the aspect-ratio-dependent switching trends are strongly affected by lithographic edge variations, additional simulations were performed using crossed-ellipse geometries with a \(\SI{5}{nm}\) rough edge. The roughened geometry was generated by replacing the analytic crossed-ellipse boundary with a binary image mask containing edge perturbations, which was then imported into MuMax3 using \texttt{ImageShape}. The same field-relaxation protocol and \(H_k\)-extraction procedure used for the smooth geometries were then applied to the roughened devices.

\begin{figure}[H]
    \centering
    \includegraphics[width=0.6\textwidth, trim=0cm 7cm 0cm 0cm, clip]{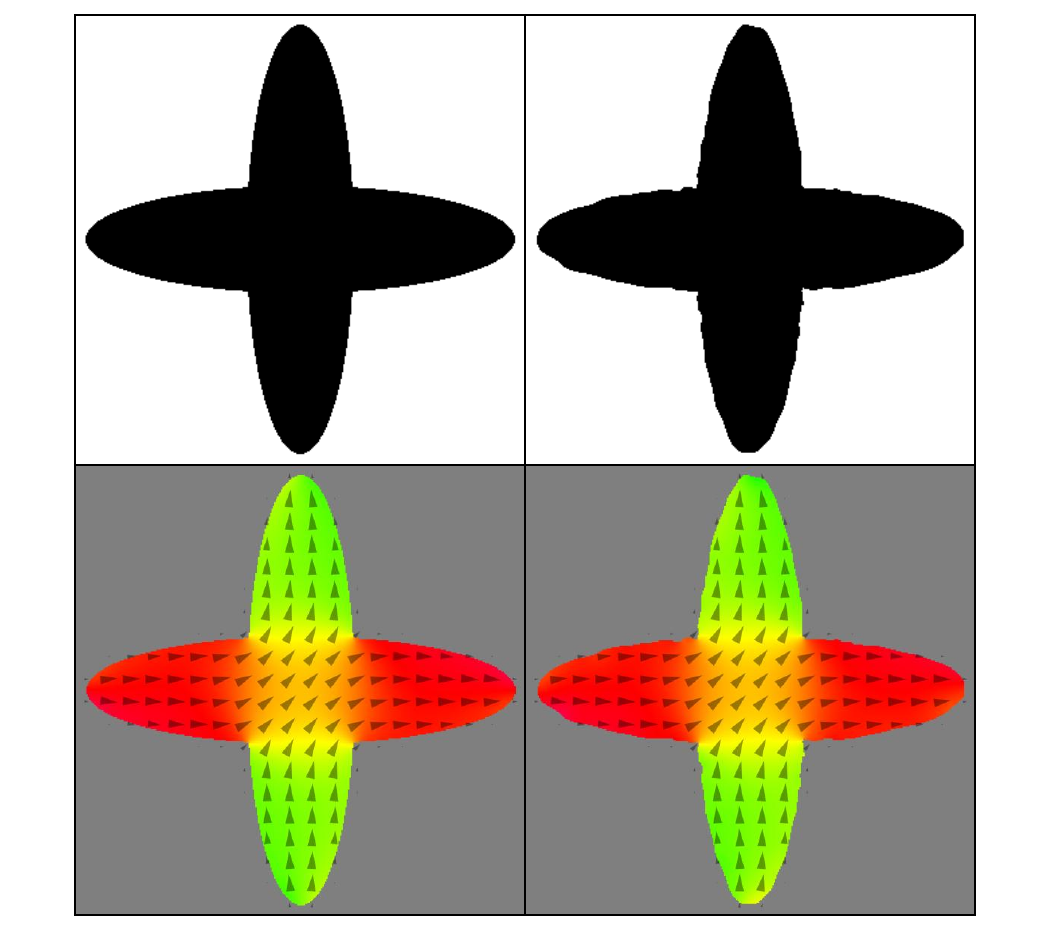}
    \caption{
    Representative smooth and roughened crossed-ellipse geometries used for the lithographic edge-roughness sensitivity test. The example corresponds to the \(a=\SI{1.6}{\micro m}\), \(a/b=4\) Py crossed ellipse. The upper panels show the binary geometry masks for the smooth geometry and the \(\SI{5}{nm}\)-rough geometry. The lower panels show the corresponding relaxed magnetic configurations. The roughened geometry preserves the same qualitative remanent-state structure as the smooth device.
    }
    \label{fig:si_rough_geometry}
\end{figure}

Figure~\ref{fig:si_rough_geometry} shows a representative comparison for the \(a=\SI{1.6}{\micro m}\), \(a/b=4\) Py crossed ellipse. The upper panels compare the smooth and roughened binary geometry masks, while the lower panels show the corresponding relaxed magnetic configurations. The roughened device retains the same qualitative remanent-state structure as the smooth crossed ellipse, indicating that the small edge perturbation does not change the basic magnetic state.

Table~\ref{tab:si_roughness_sensitivity} compares the effective anisotropy field \(H_k\) and switching field \(H_s\) for smooth and roughened devices in the \(a=\SI{1.6}{\micro m}\) Py aspect-ratio series. The \(\SI{5}{nm}\) edge roughness produces only a modest change in the switching field over the tested range. For \(a/b=8,6,\) and \(4\), \(H_s\) changes by approximately \(4\text{--}6\%\), while the extracted \(H_k\) remains close to the smooth-geometry value, especially for \(a/b=6\) and \(4\). The qualitative ordering of the aspect-ratio-dependent switching fields is therefore preserved.

\begin{table}[H]
\centering
\small
\caption{
Sensitivity of the effective anisotropy field \(H_k\) and switching field \(H_s\) to \(\SI{5}{nm}\) lithographic edge roughness for the \(a=\SI{1.6}{\micro m}\) Py crossed-ellipse series. The percentage change is calculated relative to the smooth geometry.
}
\label{tab:si_roughness_sensitivity}
\begin{tabular}{ccccccc}
\toprule
\(a/b\) &
\multicolumn{2}{c}{\(H_k\) (Oe)} &
\(\Delta H_k\) &
\multicolumn{2}{c}{\(H_s\) (Oe)} &
\(\Delta H_s\) \\
\cmidrule(lr){2-3}
\cmidrule(lr){5-6}
 & Smooth & Rough & (\%) & Smooth & Rough & (\%) \\
\midrule
8 & 32.81 & 32.16 & -1.98 & 48.4 & 46.4 & -4.13 \\
6 & 20.40 & 20.36 & -0.20 & 37.6 & 35.2 & -6.38 \\
4 &  8.09 &  8.08 & -0.12 & 23.4 & 22.2 & -5.13 \\
\bottomrule
\end{tabular}
\end{table}

These results indicate that the main aspect-ratio trend is not dominated by small edge-geometry perturbations at the \(\SI{5}{nm}\) level. This test should not be interpreted as a complete stochastic lithography study, since real patterned devices may include correlated line-edge roughness, local redeposition, edge damage, and thickness nonuniformity. Nevertheless, the comparison provides a useful robustness check showing that the smooth-geometry conclusions remain qualitatively stable under the tested roughness perturbation.

\section{\space Independent MuMax+ cross-validation}
\label{sec:si_mumaxplus_validation}

\begin{figure}[H]
    \centering
    \includegraphics[
        width=0.95\textwidth,
        trim=0cm 6cm 0cm 0cm,
        clip
    ]{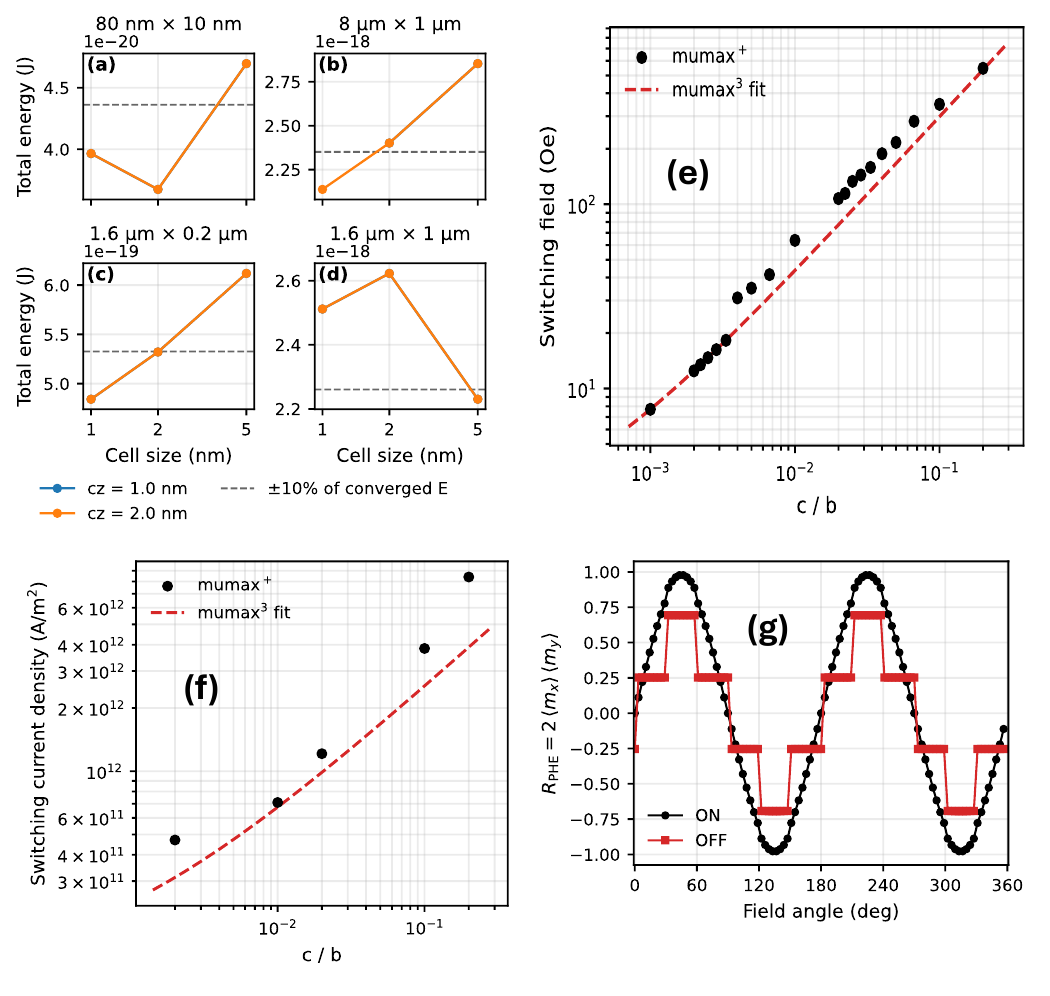}
    \caption{Independent MuMax+ cross-validation of the principal micromagnetic results.
    Panels (a)--(d) show total-energy convergence with in-plane cell size for
    representative crossed-ellipse geometries spanning the investigated size
    range; the horizontal dashed lines indicate the \(\pm10\%\) interval around
    the reference energy used in the convergence comparison.
    Panel (e) shows the fixed-\(a:b=8:1\) switching-field scaling as a function
    of \(c/b\), with the independent MuMax+ results compared with the MuMax3
    phenomenological trend.
    Panel (f) shows the corresponding switching-current-density scaling,
    demonstrating the same increase of \(J_s\) with increasing \(c/b\).
    Panel (g) shows the independently calculated angular field-on and field-off
    planar-Hall response for the
    \(1.6~\si{\micro m}\times0.8~\si{\micro m}\),
    \(a/b=2\) geometry. The field-off response resolves twelve discrete
    remanent plateaus over a full \(360^\circ\) rotation.
    Together, the calculations independently reproduce the principal
    size-scaling and multistate trends obtained using MuMax3.}
    \label{fig:si_mumaxplus_validation}
\end{figure}

To test whether the principal conclusions depend on the specific
micromagnetic implementation, selected calculations were independently
repeated using MuMax+. The cross-validation addressed four aspects of the
present study: mesh convergence, switching-field scaling at fixed
\(a:b=8:1\), switching-current-density scaling, and the low-aspect-ratio
multistate response. Representative results are summarized in
Fig.~\ref{fig:si_mumaxplus_validation}.

The mesh-convergence calculations in
Figs.~\ref{fig:si_mumaxplus_validation}(a)--(d) were performed for selected
device dimensions spanning the size range considered in the main text.
The total energy shows a clear dependence on the in-plane discretization,
whereas changing the out-of-plane cell size produces a much smaller effect
for the tested cases. An in-plane discretization of approximately
\(\SI{2}{nm}\) provides a suitable compromise between numerical accuracy and
computational cost in these independent calculations. Consistent with this
observation, the central
\(1.6~\si{\micro m}\times0.8~\si{\micro m}\) MuMax3 angular-response
calculation was repeated using a refined
\(\Delta x=\Delta y=\SI{2}{nm}\) mesh and again produced twelve discrete
field-off plateaus.

For the fixed-\(a:b=8:1\) size-scaling series, the independently calculated
switching fields are shown in
Fig.~\ref{fig:si_mumaxplus_validation}(e). The MuMax+ results follow
approximately
\[
H_s\propto(c/b)^{0.79},
\]
compared with the exponent \(n=0.85\) obtained from the MuMax3
phenomenological fit. The close agreement in the scaling exponent supports
the conclusion that the strong increase of switching field with
miniaturization is a robust geometry-dependent trend rather than a
solver-specific feature.

The current-driven cross-validation is shown in
Fig.~\ref{fig:si_mumaxplus_validation}(f). The independent calculations
reproduce the increase of switching current density with increasing
\(c/b\), with an approximate dependence
\[
J_s\propto(c/b)^{0.54}.
\]
The absolute current-density thresholds show a larger quantitative
sensitivity than the quasi-static switching fields, as expected from their
dependence on the torque implementation, discretization, pulse protocol, and
switching criterion. The comparison is therefore interpreted primarily as
validation of the current-density scaling trend.

The low-aspect-ratio multistate response was independently recalculated for
the \(a=\SI{1.6}{\micro m}\), \(b=\SI{0.8}{\micro m}\) crossed ellipse.
As shown in Fig.~\ref{fig:si_mumaxplus_validation}(g), the field-on response
remains approximately continuous, whereas the field-off response separates
into twelve discrete remanent plateaus over one full rotation. Thus, both
the refined \(\SI{2}{nm}\) MuMax3 calculation and the independent MuMax+
calculation reproduce the same twelve-state topology. This agreement
provides strong evidence that the additional low-aspect-ratio remanent
configurations arise from the crossed-ellipse energy landscape rather than
from a particular mesh or micromagnetic solver.

The independent minimum-energy-path calculations provide an additional
qualification concerning thermal stability. They confirm the weak separation
of the closest canted states and show that direct strings between
non-neighbouring endpoints can give apparently large barriers even when a
lower-energy transition route exists through intermediate remanent minima.
The physically relevant retention barrier of a state must therefore be
determined from its lowest neighbouring-state escape pathway. Consequently,
the twelve accessible plateaus are robust as a micromagnetic state-count
result, whereas determination of the number of thermally independent
nonvolatile states requires a complete neighbouring-state barrier network and
further geometry and material optimization.

\bibliographystyle{unsrt}
\bibliography{references}